\documentclass[twocolumn,showpacs,nofootinbib,prd,aps,amsmath,amssymb,superscriptaddress]{revtex4-1}
\usepackage{graphicx,epsfig}
\usepackage{bm}
\usepackage{color}
\newcommand{\beq}{\begin{equation}}
\newcommand{\eeq}{\end{equation}}
\newcommand{\ba}{\begin{eqnarray}}
\newcommand{\ea}{\end{eqnarray}}
\newcommand{\bse}{\begin{subequations}}
\newcommand{\ese}{\end{subequations}}

\newcommand{\B}{{\cal{B}}}
\newcommand{\Z}{{\cal{Z}}}
\newcommand{\M}{{\cal {M}}}

\newcommand{\DD}{{\cal {D}}}

\newcommand{\ee}{\textrm{\bf{e}}}
\newcommand{\tbb}{t_{\textrm{\tiny{bb}}}}
\newcommand{\btbb}{\bar{t}_{\textrm{\tiny{bb}}}}

\newcommand{\RR}{{}^3{\cal R}}

\newcommand{\EE}{{\cal{E}}}
\newcommand{\FF}{{\cal{F}}}

\newcommand{\HH}{{\cal{H}}}
\newcommand{\KK}{{\cal{K}}}

\newcommand{\Da}{\delta^{(A)}}

\newcommand{\Dth}{\delta^{(\Theta)}}

\newcommand{\Drho}{\delta^{(\rho)}}
\newcommand{\dDrho}{\dot{\delta}^{(\rho)}}
\newcommand{\ddDrho}{\ddot{\delta}^{(\rho)}}

\newcommand{\DKK}{\delta^{(\KK)}}

\newcommand{\DDth}{\textrm{D}^{(\Theta)}}

\newcommand{\DDthnl}{\textrm{D}_{\textrm{\tiny{NL}}}^{(\Theta)}}
\newcommand{\dDDthnl}{\dot{\textrm{D}}_{\textrm{\tiny{NL}}}^{(\Theta)}}

\newcommand{\DDthas}{\textrm{D}_{\textrm{\tiny{as}}}^{(\Theta)}}
\newcommand{\dDDthas}{\dot{\textrm{D}}_{\textrm{\tiny{as}}}^{(\Theta)}}

\newcommand{\dDDth}{\dot{\textrm{D}}^{(\Theta)}}

\newcommand{\DDKK}{\textrm{D}^{(\KK)}}

\newcommand{\DDKKnl}{\textrm{D}_{\textrm{\tiny{NL}}}^{(\KK)}}

\newcommand{\DDKKas}{\textrm{D}_{\textrm{\tiny{as}}}^{(\KK)}}
\newcommand{\DDa}{\textrm{D}^{(A)}}

\newcommand{\dd}{{\rm{d}}}

\newcommand{\Aav}{\langle A\rangle}
\newcommand{\rhoav}{\langle \rho\rangle}

\newcommand{\Thav}{\langle \Theta\rangle}
\newcommand{\KKav}{\langle \KK\rangle}

\newcommand{\Drhonl}{\delta_{\textrm{\tiny{NL}}}^{(\rho)}}
\newcommand{\dDrhonl}{\dot{\delta}_{\textrm{\tiny{NL}}}^{(\rho)}}
\newcommand{\ddDrhonl}{\ddot{\delta}_{\textrm{\tiny{NL}}}^{(\rho)}}

\newcommand{\Drhoas}{\delta_{\textrm{\tiny{as}}}^{(\rho)}}
\newcommand{\dDrhoas}{\dot{\delta}_{\textrm{\tiny{as}}}^{(\rho)}}
\newcommand{\ddDrhoas}{\ddot{\delta}_{\textrm{\tiny{as}}}^{(\rho)}}

\newcommand{\bx}{\mathbf{x}}
\newcommand{\abar}{\bar{a}}
\newcommand{\abaras}{\bar{a}_{\textrm{\tiny{as}}}}
\newcommand{\rhobar}{\bar{\rho}}
\newcommand{\rhobaras}{\bar{\rho}_{\textrm{\tiny{as}}}}
\newcommand{\rhoibaras}{\bar{\rho}_0{}_{\textrm{\tiny{as}}}}

\newcommand{\Thbar}{\bar{\Theta}}
\newcommand{\Thbaras}{\bar{\Theta}_{\textrm{\tiny{as}}}}

\newcommand{\Kbar}{\bar{\mathcal{K}}}
\newcommand{\Kbaras}{\bar{\KK}_{\textrm{\tiny{as}}}}
\newcommand{\Kibaras}{\bar{\KK}_0{}_{\textrm{\tiny{as}}}}

\newcommand{\drhocpt}{\delta_{\textrm{\tiny{CPT}}}}
\newcommand{\ddrhocpt}{\dot\delta_{\textrm{\tiny{CPT}}}}

\newcommand{\Thetacpt}{\Theta_{\textrm{\tiny{CPT}}}}
\newcommand{\dThetacpt}{\dot\Theta_{\textrm{\tiny{CPT}}}}
\newcommand{\KKcpt}{\KK_{\textrm{\tiny{CPT}}}}

\begin{document}

\title[On spherical dust fluctuations: exact vs perturbative treatment.]{On spherical dust fluctuations: the exact vs.~the perturbative approach} 

\author{ Roberto A. Sussman}
\altaffiliation{}
\affiliation{Instituto de Ciencias Nucleares, Universidad Nacional Aut\'onoma de M\'exico (ICN-UNAM),
A. P. 70--543, 04510 M\'exico D. F., M\'exico.}
\email{sussman@nucleares.unam.mx}
\author{ Juan Carlos Hidalgo}
\affiliation{Instituto de Ciencias F\'\i sicas, Universidad Nacional Aut\'onoma de M\'exico (ICF-UNAM),
A. P. 48--3, 62251 Cuernavaca, Morelos, M\'exico.}
\email{hidalgo@fis.unam.mx}
\author{Peter K. S. Dunsby}
\affiliation{Astrophysics Cosmology \& Gravity Center, and Department of Mathematics \& Applied Mathematics, University of Cape Town, 7701 Rondebosch, South Africa.}
\affiliation{ South African Astronomical Observatory, Observatory 7925, Cape Town, South Africa.}
\author{Gabriel German}
\affiliation{Instituto de Ciencias F\'\i sicas, Universidad Nacional Aut\'onoma de M\'exico (ICF-UNAM),
A. P. 48--3, 62251 Cuernavaca, Morelos, M\'exico.}
\date{\today}
\begin{abstract} 
We examine the relation between the dynamics of Lema\^\i tre--Tolman--Bondi (LTB) dust models (with and without $\Lambda$) and the dynamics of dust perturbations in two of the more familiar formalisms used in cosmology: the metric based Cosmological Perturbation Theory (CPT) and the  Covariant Gauge Invariant (GIC) perturbations.  For this purpose we recast the evolution of LTB models in terms of a covariant and gauge invariant formalism of local and non--local ``exact fluctuations'' on a {Friedmann--Lema\^{i}tre--Robertson--Walker (FLRW) background} defined by suitable averages of covariant scalars. We examine the properties of these fluctuations, which can be defined for a confined comoving domain or for an asymptotic domain extending to whole time slices. In particular, the non--local density fluctuation provides a covariant and precise definition for the notion of the ``density contrast''. We show that in their linear regime these LTB exact fluctuations (local and non--local) are fully equivalent to the conventional cosmological perturbations in the synchronous-comoving gauge of CPT and to GIC perturbations. As an immediate consequence, we show the time-invariance of the spatial curvature perturbation in a simple form. The present work may provide important theoretical connections between the exact and perturbative (linear or no--linear) approach to the dynamics of dust sources in General Relativity.
\end{abstract}

\maketitle
\section{Introduction.}

{Galaxy surveys represent a key probe of the fundamental properties of our universe. Inhomogeneities in the distribution of galaxies can be related to the underlying inhomogeneous distribution of dark matter. Consequently, by observing fluctuations in the galaxy distribution at different redshifts, one can both study the growth of dark matter perturbations and probe the nature of the gravitational action. An essential tool for describing and understanding cosmic dynamics on different scales is the study of perturbations on a FLRW background. The most favored approach to perturbations is the framework generically known as Cosmological Perturbation Theory (CPT) which relies on the smallness of quantities that describe fluctuations from the homogeneous and isotropic FLRW spacetime (see e.g. \cite{Zel:70,tomita,bardeen} for pioneering work).} This approach is based on suitably defined Gauge Invariant quantities whose 
definition and evolution equations can be found in the essential reviews
 (e.g. \cite{Mukhanov,Bernardeau,malik:wands}). While CPT is based on metric perturbations, there is an alternative and equivalent ``Gauge Invariant Covariant'' (GIC) formalism based on covariant tensorial quantities defined by a 4--velocity field \cite{BDE,1plus3,TCM}.

Perturbations based on CPT are adequate (and widely employed) in the study of 
cosmic sources {during} the early stages of evolution of the Universe, where it is safe to assume near homogeneous conditions. {This approximation is supported by the nearly isotropic (to within one part in $10^5$) Cosmic Microwave Background Radiation, which together with the {\em Almost Geren and Sachs Theorem} \cite{almost:geren} provides a strong motivation for using a spacetime close to a FLRW  model.}

At late times, however, relativistic linear perturbations based on CPT are only adequate for scales comparable to the Hubble {radius $\lambda_H$}. On scales much smaller than $\lambda_H$, the formation of cosmic structure is the dominant gravitational process. This is highly non-linear but assumed to take place in non--relativistic Newtonian conditions, so it is usually studied by means a wide range of Newtonian gravity models ranging from simple toy models (``Top hat'' or ``spherical collapse'' \cite{Padma}) to more sophisticated numerical N-body  simulations \cite{Bernardeau}. 

The study of gravitational collapse through CPT has improved by extending the scope to the non-linear regime (see e.g. \cite{Bru-Ma-Mol,tomita:nonlinear,bruni:2013,bruni:2014,rampf:rigo}). Within the perturbative approach, however, only the mildly non-linear regime can be modelled, a far from complete analysis of the collapse process up to the virilarisation stage where the density contrast is of order $\delta \gtrsim 5$.

{This leaves an important area unexplored, namely, how non-linear relativistic corrections impact on the formation of large scale structure, see for example \cite{nonlinear} for an extensive review. Indeed, some of these effects have begun to be taken into account in N-body methods, which make use of relativistic corrections to the potentials \cite{relativisticNbody1,relativisticNbody2}.}

From a non--perturbative perspective, the spherically symmetric exact solutions of Einstein's equations generically known as  Lema\^\i tre--Tolman--Bondi (LTB) dust models provide an idealized, but useful, toy model description of inhomogeneous configurations of astrophysical and cosmological interest (see comprehensive reviews of these models in \cite{kras1,kras2,BKHC2009}). While a nonzero $\Lambda$ term can be easily incorporated into the dynamics of these exact solutions, these solutions have been widely used to model large scale CDM density voids to fit observational data without assuming the existence of dark energy or a cosmological constant (see reviews in \cite{bisnotwal,marranot}). Moreover, if we assume that $\Lambda>0$, LTB models become an inhomogeneous generalization of the $\Lambda$--CDM model describing exact non--perturbative CDM inhomogeneities in a $\Lambda$--CDM background favored by observations (see \cite{Romano}). In fact, observational data also fit LTB models with $\Lambda>0$ and an FLRW background that is not necessarily the usual $\Lambda$--CDM background \cite{LLTB}.

Introducing a representation based on covariant scalars (q--scalars \cite{part1}) and their associated  ``exact fluctuations'' allows for a clear study of important properties of LTB models: their phase space evolution as a dynamical system \cite{sussDS2,sussmodes}, their radial asymptotics \cite{RadAs}, the nature and evolution of density profiles \cite{RadProfs}, as well as their use to probe theoretical formalisms of space-time averaging \cite{sussBR} (see review in \cite{part1}) and gravitational entropy \cite{susslar}. It is important to remark that in these references the exact fluctuations were called ``exact perturbations'', which may not be a convenient name because the term ``perturbation'' is commonly understood to refer to approximated (not exact) quantities.

It is a well known fact (see extensive work in \cite{part2}) that the q--scalars and their fluctuations, in their local and non--local versions, fully determine the dynamics of LTB models recast in terms of evolution equations, analogous to those of linear perturbations on an FLRW background.  This resemblance can be reframed in precise unambiguous terms by a rigorous correspondence maps that give rise to a rigorous covariant and gauge invariant perturbation formalism. In particular, it can be shown that the density fluctuation can be expressed in terms of exact covariant expressions that generalize the density growing and decaying modes of linear dust perturbations~\cite{sussmodes}.  Also, the non--local density fluctuation provides a precise covariant characterization of the intuitive notion of the ``density contrast'', a concept loosely, and often incorrectly, employed in many astrophysical and cosmological applications of LTB models.   

In the present paper we extend the above-mentioned studies by establishing equivalences between the perturbative CPT and GIC quantities and exact inhomogeneities defined through the exact fluctuations. Throughout this paper, we are careful to stress the fact that the evolution equations for the fluctuations do not describe ``small''  deviations from a FLRW background, but the evolution of exact quantities of an exact solution of GR (LTB models). Yet we show how, in a suitable linear regime, these fluctuations reduce to the spherically symmetric linear perturbations of the GIC and CPT formalisms. This result is summarised in Table~I of Sec.~\ref{sec:summary}.  In verifying this correspondence we consider the comoving gauge of CPT dust perturbations (as LTB models are defined in a comoving frame). We argue that the exact fluctuations represent a generalisation of the usual perturbation scalars to the non-linear regime, as first suggested in \cite{part2} and here extended to the case $\Lambda>0$. Analysing such generalisations is important to determine the fate of small fluctuations throughout the non-linear stages of structure formation, a regime poorly explored in relativistic cosmology. We also show that the time conservation of the spatial curvature perturbation of CPT theory can be expressed (up to linear terms) in terms of time preserved quantities of LTB models.    

The paper is organized as follows. In section \ref{LTB-dust} we introduce the LTB models in terms of the q--scalars formed from the standard fluid flow covariant LTB scalars: the energy density $\rho$, the Hubble expansion
$\Theta$, and the spatial curvature $\KK=\RR/6$ (where $\RR$ is the three-dimensional Ricci scalar). In section~\ref{QL-Perts} we define the fluctuations as exact deviations between the q--scalars $A_q$ and the standard covariant scalars $A$. In section~\ref{NL-Perts} we introduce the non--local fluctuations as exact deviations with respect to the ``q--averages'', which are the non--local functionals associated with the q--scalars $A_q$. Asymptotic non--local fluctuations are discussed in section~\ref{Exact-Perts}. The conditions that define a linear regime in LTB exact fluctuations are given in section~\ref{Linear-Regime}. The comparison between all fluctuations in the linear regime with the CPT formalism is described in section~\ref{sec:CPT}, while the correspondence with the GIC perturbations is discussed in section~\ref{CGI-Perts}. We summarise and discuss our results in the final Section~\ref{sec:summary}, where we present a useful perturbation-to-fluctuation dictionary in Table~I. The relation between the LTB metric variables that we used and the standard ones is given in Appendix~\ref{app:a}. We examine in Appendix~\ref{app:b} the Darmois matching conditions that are used for the rigorous definition of an FLRW background for the exact fluctuations, while the form of LTB metric functions in the linear regime are discussed in Appendix~\ref{app:c}.   
\noindent
\section{LTB dust models in the q--scalar representation.}
\label{LTB-dust}
A convenient parametrization of LTB dust models is given by the following useful FLRW--like metric (the relation with the standard metric variables is given in Appendix~\ref{app:a}):  
\beq 
  \dd s^2 =-\dd t^2+ a^2\left[\frac{\Gamma^2}{1-\KK_{q0} r^2}\dd
  r^2+r^2\left(\dd\theta^2+\sin^2\theta\dd\varphi^2\right)\right],
\label{ltb2}
\eeq 
where the scale factors $a=a(t,r)$ and $\Gamma=\Gamma(t,r)$ satisfy:
\ba
\dot a^2 &=& \frac{8\pi}{3}\frac{\rho_{q0} }{a}-\KK_{q0} +\frac{8\pi}{3}\Lambda\,a^2,\quad \mathrm{with}\; \dot a =\frac{\partial a}{\partial t},\qquad \label{Friedman}\\
\Gamma &=& 1+\frac{r a'}{a},\qquad \qquad\qquad \qquad  \mathrm{with}\;   a' =\frac{\partial a}{\partial r},\label{Gamma}
\ea
while the functions $\KK_{q0}(r) $ and $\rho_{q0}(r)$ are defined further ahead (see Eq.~\eqref{rhoKKHH}). The subindex ${}_0$ will denote henceforth evaluation at an arbitrary fiducial hypersurface $t=t_0$, which can be  taken as the present cosmic time. Notice that we have chosen the radial coordinate so that $a_0=\Gamma_0=1$.

The standard approach to LTB models is based on using the solutions (whether analytic or numerical) of \eqref{Friedman} to determine the metric functions $a$ and $\Gamma$ in order to compute all relevant quantities. We follow here a different approach, based on {a set of} useful alternative variables called ``q--scalars'',  constructed with the standard covariant scalars  \cite{sussmodes,part1,part2} 
\footnote{The connection between these integral definitions and a weighted proper volume average is discussed in section \ref{NL-Perts}. See a comprehensive discussion in \cite{part1,sussBR}.} 
\beq A_q =\frac{\int_0^r{A\,R^2\,R'\,\dd\bar r}}{\int_0^r{\,R^2\,R'\,\dd\bar r}}=\frac{3\int_0^r{A\,R^2\,R'\,\dd\bar r}}{R^3},\quad A=\rho,\,\Theta,\,\KK,\label{Aqdef} \eeq
where $R=a r$ and $\rho$ is the energy
density. The homogeneous expansion is  $\Theta= \bar \nabla_au^a$, with $\bar\nabla$ the gradient projected in the hypersurfaces orthogonal to $u^a$. Also $\,\KK=\RR/6$ is the spatial curvature of these hypersurfaces, with $\RR$ the three-Ricci scalar. The scalars $A$ and $A_q$ in
Eq.~\eqref{Aqdef} are related through the following ``exact fluctuations" 
\footnote{We discuss in detail the notion of an ``exact fluctuation"
  in the following section. The q--scalars and the exact fluctuations
  are directly related to curvature and kinematic scalars \cite{part1}. The domain of integration in the integrals in
  \eqref{Aqdef} and \eqref{Drho}--\eqref{DDKK} is a spherical comoving
  domain $\DD[r]$ parametrized by $0\leq \bar r\leq r$, where $\bar r
  = 0$ is a symmetry center. See \cite{part1,part2,sussmodes} for a comprehensive
  discussion on the definition and properties of these variables.}  
\bse\ba  \Drho\equiv&&\, \frac{\rho-\rho_q}{\rho_q} =
\frac{r\rho'_q/\rho_q}{3\Gamma} =
\frac{1}{\rho_q\,R^3}\int_0^r{\rho'\,R^3\dd \bar r},\label{Drho}\\ 
 \DDth \equiv&&\,
 \Theta-\Theta_q=\frac{r\Theta'_q}{3\Gamma}=\frac{1}{R^3}\int_0^r{\Theta'\,R^3\dd
   \bar r},\label{DDh}\\  
 \DDKK \equiv&&\,
 \KK-\KK_q=\frac{r\KK'_q}{3\Gamma}=\frac{1}{R^3}\int_0^r{\KK'\,R^3\dd
   \bar r}.\label{DDKK} 
\ea\ese 
The q--scalars and the exact fluctuations $\Drho,\,\DDth$ and $\DDKK$
satisfy the following scaling laws 
derived from the energy conservation equation and the $ij$ components
of the Einstein equations \cite{part1,part2,sussmodes}:
\ba  \rho_q =\frac{\rho_{q0} }{a^3}, \qquad \KK_q=\frac{\KK_{q0} }{a^2},\qquad \frac{\Theta_q}{3}=\frac{\dot a}{a},\label{rhoKKHH}\\
 1+\Drho= \frac{1+\Drho_0}{\Gamma},\qquad \frac{2}{3}+\DKK = \frac{2/3+\DKK_0}{\Gamma}.\label{DrhoKK}\ea
These are complemented by the algebraic constraints (analogous to the ``Hamiltonian'' and spatial curvature constraints)
 \ba 
\left(\frac{\Theta_q}{3}\right)^2 = 
\frac{8\pi}{3}(\rho_q+\Lambda) -\KK_q,\label{HHq}\\
\frac{3}{2}\DDKK = 4\pi\rho_q\Drho - \frac{1}{3}\Theta_q\DDth,\label{DDH}
\ea
where the subindex ${}_0$ denotes evaluation at an arbitrary fixed
$t=t_0$ and we have introduced, together with $\Drho$, in \eqref{DrhoKK}, the relative exact
fluctuation   
\footnote{The term ``perturbation'' was used in
  \cite{part1,part2,sussmodes} only to denote the dimensionless
  quotient fluctuations  $\Drho,\,\Dth,\,\DKK$, while $\DDth$ and
  $\DDKK$ in \eqref{DDh} and \eqref{DDKK} were called
  ``fluctuations''. In this article the term ``exact fluctuations'' will
  denote both the $\Da$ and the $\DDa$. We consider as basic set of
  exact fluctuations the quantities $\{\Drho,\,\DDth,\,\DDKK\}$ because
  they provide a straightforward link to perturbation formalisms in
  the literature in which only the density perturbation is
  constructed in the dimensionless quotient form \eqref{Drho}
  (inspired on the intuitive notion of the density
  contrast). Besides this point, the exact relative fluctuations
  $\Dth$ and $\DKK$ constructed as in \eqref{DKK}can become ill--defined (they
  diverge) if $\Theta_q$ or $\KK_q$ (which appear in the denominator)
  vanish, which can occur in physically interesting scenarios in LTB
  models, for example: $\Theta_q=0$ occurs at the ``bounce'' from
  expansion to collapse in collapsing models, or $\KK_q=0$ necessarily
  holds along a comoving ``boundary'' layer separating comoving
  regions in which  $\KK_q$ switches sign.}: 
\ba 
 \DKK = \frac{\DDKK}{\KK_q}=\frac{\KK-\KK_q}{\KK_q}.\label{DKK} 
\ea  
Any LTB model becomes fully determined, either analytically (if $\Lambda=0$ or in certain cases with $\Lambda>0$ \cite{sussDS2}) or numerically (the general case $\Lambda>0$), and can be uniquely specified by selecting a value of $\Lambda$ and, as free parameters or initial conditions, any two of the initial value functions  $\{\rho_{q0} ,\,\KK_{q0} ,\,\Theta_{q0} \}$. 

The analytic forms \eqref{rhoKKHH}--\eqref{DKK} are exact solutions of the evolution equations constructed from the variables $\rho_q,\,\Theta_q$ and $\Drho,\,\DDth$ \cite{sussmodes,part2},
\bse\ba  
\dot\rho_q &=& -\rho_q\Theta_q,\label{FFq1}\\
 \dot \Theta_q &=& -\frac{\Theta_q^2}{3}-4\pi\rho_q+8\pi\Lambda, \label{FFq2}\\
 \dDrho &=& -(1+\Drho)\,\DDth,\label{FFq3}\\
  {\dDDth} &=&  -\left(\frac{2}{3}\Theta_q +\DDth\right)\DDth- {4\pi}\rho_q\Drho,\label{FFq4}
\ea\ese
subject to the algebraic constraints \eqref{HHq}--\eqref{DDH}, which
will hold for all $t$ once we solve them by specifying initial
conditions at arbitrary $t=t_0$. Combining the evolution equations \eqref{FFq1}--\eqref{FFq2} leads to the second order equation
\beq \ddDrho -\frac{2\left[\dDrho\right]^2}{1+\Drho}+\frac{2}{3}\Theta_q\dDrho-4\pi\rho_q\Drho(1+\Drho)=0,\label{secord1}\eeq 
which is an exact generalization of the well known evolution equation of linear dust perturbations in the comoving gauge \cite{lyth:liddle}. 
The constraints \eqref{HHq}--\eqref{DDH} allow for the construction of systems equivalent to \eqref{FFq1}--\eqref{FFq4}, but based on alternative set of variables $A_q$ and/or relative fluctuations $\Da$ (see examples for the case $\Lambda=0$ in
equations (21a)--(21d)  of \cite{part2}).  

\section{Local exact fluctuations.}
\label{QL-Perts}
It is intuitively clear that   we can identify in the system
\eqref{FFq1}--\eqref{FFq4} the subset of evolution equations
\eqref{FFq1}--\eqref{FFq2} for FLRW--like ``background variables''
$\rho_q,\,\Theta_q,\,\Lambda$, as these are identical to FLRW evolution
equations for their equivalent FLRW scalars (although $q-$scalars also
carry a spatial dependence). On the other hand the subset
\eqref{FFq3}--\eqref{FFq4} corresponds to the evolution equations of
the exact fluctuations defined in \eqref{Drho}--\eqref{DDKK}.  

\subsection{The notion of an ``exact fluctuation''.}

The connection between  ``exact fluctuations'' and ``perturbations'' requires further clarification given the common use of these terms in the
literature. Consider for example the relation $\rho=\rho_q(1+\Drho)$
that follows from \eqref{Drho}: this is an exact relation, and thus it
 does not require a small parameter expansion to describe
departures of $\rho$ from $\rho_q$, since both $\rho$ and $\rho_q$ are exact LTB
scalars (in other words: we have not assumed and {\it need not assume} a small $|\Drho|$). The same
argument goes for the relations between $\Theta$ vs $\Theta_q$ and $\KK$ vs
$\KK_q$ that follow from \eqref{DDh} and \eqref{DDKK}. Hence, we call
$\Drho,\,\DDth,\,\DDKK$ ``exact fluctuations'' in order to distinguish
them from the common usage of the term ``perturbations'' in standard
formalisms of Cosmological Perturbation Theory, namely: quantities
defining a ``perturbed'' spacetime $\M$ that is ``almost  FLRW'', meaning 
that it represents a ``small departure''
from a suitable known background FLRW spacetime $\bar \M$ through a
linearization procedure applied to characteristic quantities (metric,
scalars, vectors, 
tensors) of the latter. In our case the ``exact fluctuations'' relate
an exact LTB model (the spacetime $\M$) to a precise exact FLRW 
background $\bar \M$ defined by suitable scalars (the $A_q$) of the same LTB model once we choose a given comoving domain $\DD[r]$. In other words, the $A_q$ evolve in an identical way to their FLRW equivalents when evaluated at a specific coordinate $r$.


\subsection{Fluctuation-to-perturbation correspondence maps and gauge invariance.}

\begin{figure}
\begin{center}
\includegraphics[scale=0.4]{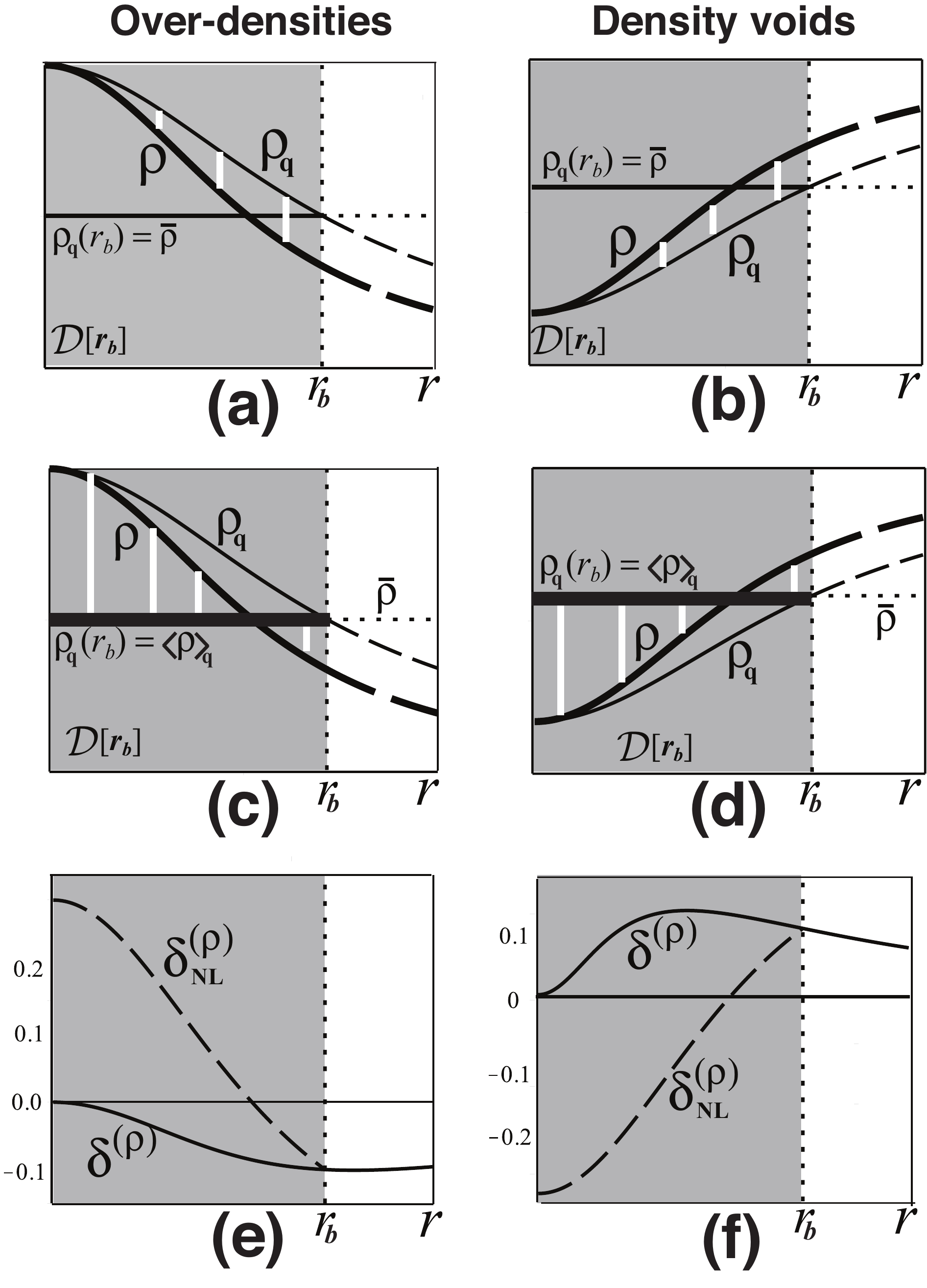}
\caption{{\bf Confined LTB local and non--local exact density fluctuations.} The panels (a)--(d) describe generic radial density profiles (over--densities and density voids) while (e) and (f) depict local and non--local density fluctuations, $\Drho$ and $\Drhonl$, confined to a comoving domain $\DD[r_b]$ marked by  $0\leq r < r_b$. Panels (a) and (b) illustrate how $\Drho$ follows from the local comparison of $\rho$ and $\rho_q$ at each $r$ in the domain, whereas panels (c) and (d) show how $\Drhonl$ follows from comparing $\rho$ at each $r$ with the q--average $\rhoav_q[r_b]$ that characterizes the whole domain. Notice that the FLRW ``background'' density is given for every domain by $\bar\rho=\rho_q(r_b)=\rhoav_q[r_b]$ and that these fluctuations can be well defined without assuming a matching with an actual FLRW spacetime.}
\label{FIG1}
\end{center}
\end{figure}

The set of exact fluctuations $\Drho,\,\DDth,\,\DDKK$ in \eqref{Aqdef} and \eqref{Drho}--\eqref{DDKK} are covariant local objects, as they provide the  
exact deviation between the covariant scalars $A= (\rho,\,\Theta,\,\KK)$ and their
corresponding q--scalars $A_q$ (which are also covariant \cite{sussmodes,part2}) along every 
concentric 2--sphere labeled by constant $r$ that marks
the boundary $\B[r]$ of an integration domain $\DD[r]$ (a spherical comoving
region). This is illustrated in Figures~\ref{FIG1} and~\ref{FIG2}. We remark that the definition of exact fluctuations can easily be extended to the non--spherical Szekeres models~\cite{sussbol}. Their role as
exact fluctuations can be defined rigorously through a covariant
and gauge-invariant formalism (see \cite{part2}).  

Since any LTB model ($\M$) and any dust FLRW spacetime ($\bar \M$)
share the same comoving geodesic 4--velocity, spherical comoving
coordinates and dust source, the appropriate correspondence mapping
that defines the exact fluctuations is furnished rigorously  by 
associating to each comoving domain
$\DD[r]$ of the LTB model
($\M$) the unique FLRW dust spacetime $\bar \M$ defined by the continuity of the 3--metric and extrinsic curvature of the common ``interface'' hypersurface $\B[r]$ (world--tube generated by comoving observers at fixed arbitrary $r$)\footnote{In previous papers, e.g. \cite{part2}, this mapping was denoted by a "perturbation" mapping, but we prefer to call it fluctuation--to--perturbation mapping to avoid the semantic problem emanating from the fact that the term "perturbation" is used to describe approximate quantities.}.
 As shown in Appendix~\ref{app:b}, this is equivalent to the conditions for a smooth match of $\M$ and $\bar \M$ at an arbitrary $\B[r]$, which implies the continuity of the q--scalars $A_q$ and the FLRW scalars $\bar  A$ at $\B[r]$ for all $t$ 
\beq 
 [\rho_q]_r=\rhobar(t),\qquad [\Theta_q]_r=\Thbar(t),\qquad [\KK_q]_r=\Kbar(t),\label{darmois}\eeq
where $[\,\,]_r$ denotes evaluation at fixed $r$.
$\{\rho_q,\,\Theta_q,\,\KK_q\}$ are given by Eq.~\eqref{rhoKKHH} and
an over bar will hereafter denote FLRW scalars.  It is
important to remark that this identification of $\M$ and $\bar \M$ is strictly a rigorous and precise procedure to define an FLRW background and q--perturbations for every $\DD[r]$ of a generic LTB model: it {\it does not} require that we undertake an actual matching of the domain $\DD[r]$ and $\bar \M$ in the form of a ``Swiss Cheese'' configuration (see Figures~\ref{FIG1} and~\ref{FIG2} and reference \cite{part2} for a comprehensive discussion).

The gauge invariance (GI) of the exact fluctuations follows from the Stewart--Walker lemma \cite{Stewart:1974uz}: a GI ``perturbation'' is any
nonzero quantity in $\M$ that vanishes in the background $\bar \M$ for
a given perturbation formalism in which $\M$ and $\bar  \M$ have been
defined and related through suitable mappings (as for example the map
introduced in \cite{part2} summarized above). It is important to bear
in mind that the conditions in Eq.~\eqref{darmois} do not involve the continuity
of the usual covariant scalars $A$, as $[A]_r\ne [A_q]_r$ and thus
$[\Da]_r\ne 0$ and $[A]_r\ne \bar  A(t)$ hold in general for an arbitrary
$r$ \cite{part2} (see Figure~\ref{FIG2} and Appendix~\ref{app:b}). As a consequence, the GI criterion based on Stewart's lemma
does not require $\Drho,\,\DDth,\,\DDKK$ to vanish at any fixed
$r$, but to vanish in the FLRW spacetime characterized by the scalars
$\bar  A$ in  that has been mapped through the Darmois
conditions \eqref{darmois}. In particular, if we impose as supplementary conditions besides \eqref{darmois}, that
$[\Drho]_r=[\DDth]_r=[\DDKK]_r=0$ hold for any given fixed finite $r$; then
we are forcing $[A]_r=[A_q]_r=\bar  A$. This leads to ``Swiss
Cheese'' models of exact fluctuations (see Figure~\ref{FIG2} and Appendix~\ref{app:b}). 
\section{Non--local exact fluctuations.} 
\label{NL-Perts}

We can also consider \eqref{Aqdef} as the correspondence rule of a linear functional for a
single (but arbitrary) domain $\DD[r_b]$. This functional is the
``q--average'', which assigns to each scalar $A$ and each arbitrary
fixed comoving domain $\DD[r_b]$ the real number~\cite{part1,part2} 
\footnote{The q--average $\Aav_q[r_b]$ is the proper volume average of $A$ with weight factor $\FF=\sqrt{1-\KK_{q0} r^2}$ over the comoving domain $\DD[r_b]$ with $0\leq r<r_b$. A detailed comparison with the standard proper volume average emerging from Buchert's formalism is given in \cite{part1,part2} (see also \cite{sussBR}).}
\beq
(A,\DD[r_b])\mapsto \Aav_q[r_b](t)=\frac{\int_0^{r_b}{A R^2 R' \dd\bar
      r}}{\int_0^{r_b}{R^2 R' \dd\bar r}},\label{q-ave}
\eeq
where $R= ar$. It is important to remark that the q--averages and the
q--scalars $A_q$ are different objects: the q--averages are non--local because 
 $\Aav_q[r_b]$ is a single real
number assigned to the whole domain $\DD[r_b]$, and thus must be
treated as an effective constant for inner concentric domains $\DD[r]$
with $0\leq r\leq r_b$ (see Figures \ref{FIG1}c, \ref{FIG1}d, \ref{FIG3}a and \ref{FIG3}d), whereas $A_q(t,r)$ is a function of the domain boundary and thus it smoothly varies for these inner
domains (see Figure~\ref{FIG1}). Hence, they only coincide at the domain boundary:
$A_q(t,r_b)=\Aav_q[r_b](t)$ of every $\DD[r_b]$ (see also comprehensive discussion on this in
\cite{part1,part2}).       

Since the average $\Aav_q[r_b]$ is a non--local quantity, we can construct non--local exact fluctuations in an analogous way as the exact fluctuations
$\Drho,\,\DDth$ and $\DDKK$ in \eqref{Drho}--\eqref{DDKK} \cite{part2} as follows: 
\ba   \Drhonl &=&\frac{\rho( r)-\rhoav_q[r_b]}{\rhoav_q[r_b]},\qquad \qquad \nonumber\\
 \DDthnl &=&\Theta( r) - \Thav_q[r_b],\qquad \nonumber\\
 \DDKKnl &=&\KK( r) - \KKav_q[r_b],
  \label{nl1}
\ea
such that
%
\ba
  \rho&=&\rhoav_q[r_b]\,\left[1+\Drhonl\right],\qquad \qquad \nonumber\\ 
\Theta&=&\Thav_q[r_b]+\DDthnl,\nonumber\\
 \KK &=&\KKav_q[r_b]+\DDKKnl,
\label{nl2}
\ea
where $\Drhonl,\,\DDthnl$ and $\DDKKnl$ depend on
$(t,r,r_b)$. The relations involving the gradients of $\rho,\,\Theta,\,\KK$ given by
\eqref{Drho}--\eqref{DDKK} for $\Drho,\,\DDth,\,\DDKK$ are only valid
for $r=r_b$. The non--local nature of the exact fluctuations
\eqref{nl1}  follows from the fact
 that they compare (for all $t$) 
the local values $A_q( r),\,\,0\leq r\leq r_b$ with a
non--local quantity $\Aav_q[r_b]$ assigned by Eq.~\eqref{q-ave} to the
whole domain $\DD[r_b]$ (see Figure~\ref{FIG2}).  

\subsection{Evolution equations and background variables as averages.} 

Combining \eqref{nl1} and \eqref{nl2} we obtain the relation between exact fluctuations and their non--local analogues:
\bse\ba    
\Drhonl &=&\frac{\rho_q}{\rhoav_q[r_b]}\left[1+\Drho\right]-1,\label{relperts1}
\\  
\DDthnl&=&\DDth+\Theta_q-\Thav_q[r_b],\label{relperts3}\\
\DDKKnl&=&\DDKK+\KK_q-\KKav_q[r_b],\qquad\qquad\label{relperts2}
\ea\ese
which upon substitution in \eqref{FFq1}--\eqref{FFq4} yields an analogous set of evolution equations: 
%
\bse
\ba 
  \rhoav\dot{}_q &=& -\rhoav_q\,\Thav_q,\label{NL1a}\\
  \Thav\dot{}_q &=& -\frac{1}{3}\Thav_q^2- 4\pi\rhoav_q +
8\pi\Lambda, \label{NL1b}\\ 
  \dDrhonl &=& -\left[1+\Drhonl\right]\,\DDthnl,\label{NL1c}\\
  \dDDthnl &=& -\left(2\Thav_q-\frac{4}{3}\Theta_q+\DDthnl\right)\DDthnl -\notag\\
&&\frac{2}{3}(\Theta_q-\Thav_q)^2-4\pi\rhoav_q\,\Drhonl, \label{NL1d}  
\ea
\ese
where we omitted the domain indicator in the q--averages  to simplify notation. The system \eqref{NL1a}--\eqref{NL1d} must be supplemented by the evolution equations \eqref{FFq1}--\eqref{FFq2}, since $\Theta_q$ appears explicitly in \eqref{NL1d}, and by the algebraic constraints
\bse
\ba
   \frac{\Thav_q^2}{9}&=&\frac{8\pi}{3}(\rhoav_q+\Lambda)-\KKav_q,\label{constrNL1}\\
\frac32 \DDKKnl &=&{4\pi}\rhoav_q\Drhonl-\frac{1}{3}\Thav_q\DDthnl,\label{constrNL2}
\ea
\ese
which are analogous to \eqref{HHq} and \eqref{DDH}. It is straightfoward to derive, from \eqref{NL1a}--\eqref{NL1d}, a second order equation for $\Drhonl$:
\ba   
&& \ddDrhonl -\frac{\left[\dDrhonl\right]^2}{1+\Drhonl}+\left[2\Thav_q-\frac{4}{3}\Theta_q\right]\dDrhonl -\notag\\
&&\quad\left[4\pi \rhoav_q\Drhonl -2(\Theta_q-\Thav_q)^2\right](1+\Drhonl)=0,\quad
\label{secord2}
\ea
\noindent which is analogous to Eq.~\eqref{secord1}. Notice that \eqref{NL1a}--\eqref{NL1d}, \eqref{constrNL1}--\eqref{constrNL2} and \eqref{secord2} reduce to \eqref{FFq1}--\eqref{FFq4}, \eqref{HHq}--\eqref{DDH} and \eqref{secord1} at the domain boundary $r=r_b$ for which $\rhoav_q,\,\Thav_q$ and $\rho_q,\,\Theta_q$ exactly coincide and thus $\Drhonl=\Drho$ and $\DDthnl=\DDth$ hold for all $t$. The fact that \eqref{NL1a}--\eqref{NL1b} (which involve averages) are formally the same evolution equations as \eqref{FFq1}--\eqref{FFq2} follows from the fact that back--reaction vanishes for the q--average (see \cite{part1}).  
 
As with the evolution equations \eqref{FFq1}--\eqref{FFq4} for
local exact fluctuations, we can also identify in \eqref{NL1a}--\eqref{NL1d}
the subset of FLRW--like evolution equations
\eqref{NL1a}--\eqref{NL1b} for the background variables
$\rhoav_q,\,\Thav_q$ and the subset \eqref{NL1c}--\eqref{NL1d} of
evolution equations for the exact (now non--local) fluctuations
$\Drhonl,\DDthnl$. Hence, these variables also give rise to a covariant and
gauge invariant perturbation formalism (see \cite{part2}) that is
analogous to that of the local fluctuations. Notice that
$\Aav_q[r_b]=A_q(r_b)$ holds for every $r_b$ (see Figures~\ref{FIG1} and~\ref{FIG2}), and hence the Darmois matching conditions in Eq.~\eqref{darmois} now identify an FLRW background spacetime
through the q--average of covariant scalars over domains $\DD[r_b]$.  

\subsection{The density contrast.} 

\begin{figure}
\begin{center}
\includegraphics[scale=0.4]{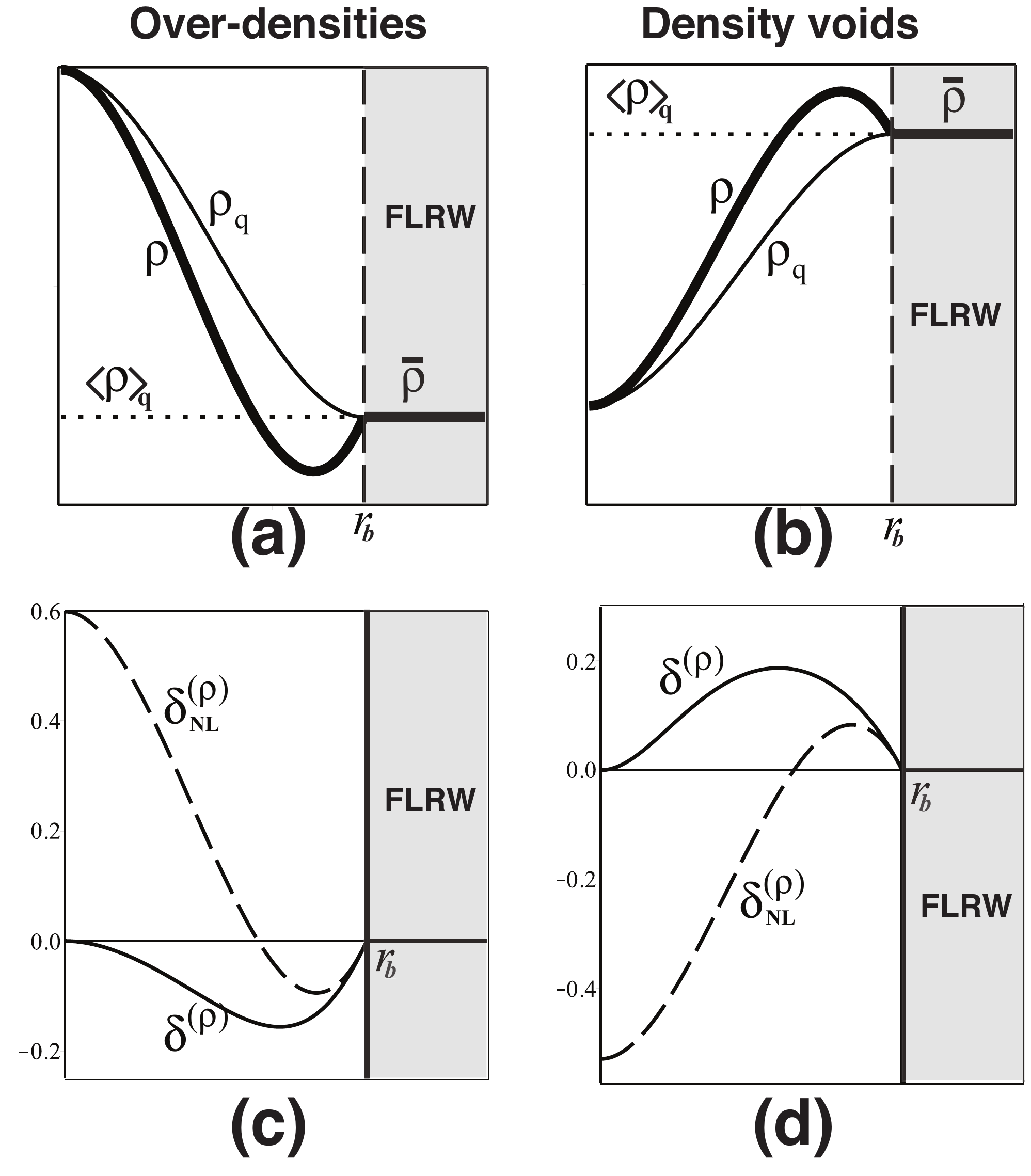}
\caption{{\bf Confined local and non--local exact density fluctuations in a Swiss Cheese model.} The panels (a)--(b) describe radial density profiles (over--density and density void) of a Swiss Cheese model made by a smooth matching of a comoving domain $\DD[r_b]$ in the radial range  $0\leq r < r_b$ of an LTB model to an FLRW spacetime. Panels (c) and (d) depict the corresponding local and non--local density fluctuations. Notice that Darmois matching conditions of Eq.~\eqref{darmois} only require continuity of $\rho_q$, so that background density is defined by $\rho_q(r_b) =\rhoav_q[r_b] =\bar \rho$. Demanding also continuity of $\rho$ and the vanishing of fluctuations at the FLRW boundary leads to the ``humps'' and ``troughs'' in the radial profiles.}
\label{FIG2}
\end{center}
\end{figure}

It is worth recalling that the non--local exact density fluctuation $\Drhonl$ defined in \eqref{nl1}--\eqref{nl2} provides a rigorous and covariant (and GI) definition for the  ``density contrast'' in a domain $\DD[r_b]$, as it compares the local density $\rho$ at each point with the FLRW background density identified by the q--average of the density $\rhoav_q[r_b]$ in this domain (see Figures~\ref{FIG1} and~\ref{FIG2}).  Therefore, equations \eqref{NL1a}--\eqref{NL1d}, as well as \eqref{secord2}, provide the evolution of the exact, non--perturbative, density contrast. Notice, however, that the sign of $\Drhonl$ is opposite to that of $\Drho$ for a given density profile: 
\begin{itemize}
\item {\bf Over--density profile}: we have $\Drhonl>0$ (positive density contrast) and $\Drho<0$ (negative gradient of $\rho_q$). See Figures~\ref{FIG1}a, \ref{FIG1}c, \ref{FIG1}e, \ref{FIG2}a, \ref{FIG2}c, \ref{FIG3}a and \ref{FIG3}c.   
\item {\bf Density void profile}: we have $\Drhonl<0$ (negative density contrast) and $\Drho>0$ (positive gradient of $\rho_q$). See Figures~\ref{FIG1}b, \ref{FIG1}d, \ref{FIG1}f, \ref{FIG2}b, \ref{FIG2}d, \ref{FIG3}b and \ref{FIG3}d.
\end{itemize}
This sign difference follows from the fact that $\Drhonl$ compares $\rho$ with $\rhoav_q[r_b]$, which remains fixed inside $\DD[r_b]$, whereas $\Drho$ is proportional to the gradient $\rho'_q$ (from \eqref{Drho}) and compares $\rho$ and $\rho_q$, which are both varying at inner points, these differences are clearly displayed in Figures~\ref{FIG1}, \ref{FIG2} and \ref{FIG3}. 
\section{Exact fluctuations on an asymptotic FLRW background.}
\label{Exact-Perts}
We have considered so far exact fluctuations (local and non--local) that are ``confined'' in bounded concentric comoving domains (Figures~\ref{FIG1} and~\ref{FIG2}).  For LTB models converging in the asymptotic radial direction to FLRW models, the metric functions and covariant scalars take the following forms as $r\to \infty$
\footnote{The conditions for LTB models to be asymptotic to a FLRW background spacetime in the spacelike radial direction at every time slice were discussed in \cite{RadAs}.  The radial rays in time slices orthogonal to $u^a$ are spacelike geodesics of the LTB metric, hence for a well defined radial coordinate the proper length along these curves is a monotonic function of $r$, and thus the proper radial asymptotic limit of any scalar is given by $r\to\infty$. Only the case $\Lambda=0$ was examined in \cite{RadAs}, but the results easily extend to the case $\Lambda>0$. If $\Lambda=0$ LTB models can be radially asymptotic to FLRW models that are spatially flat (Einstein de Sitter $\bar \KK=0$) or with negative spatial curvature (open FLRW $\bar \KK<0$), as ``closed'' FLRW lack an asymptotic radial range. If $\Lambda>0$ then convergence to a FLRW model with $\bar \KK>0$ is possible (see \cite{sussDS2}).} 
\bse\ba a\to \abaras,\qquad \Gamma\to 1,\label{FLRW1a}\\
(\rho,\,\Theta,\,\KK) \to (\rhobaras,\,\Thbaras,\,\Kbaras),\label{FLRW1b}\\
 (\Drho,\,\DDth,\,\DDKK)\to 0,\label{FLRW1c}\\
  (\rho_q,\,\Theta_q,\,\KK_q)\to (\rhobaras,\,\Thbaras,\,\Kbaras),\label{FLRW1d}\ea\ese
where $\abaras=\abaras(t),\,\rhobaras=\rhobaras(t),\,\Thbaras=\Thbaras(t)$ and $\Kbaras=\Kbaras(t)$ are the scale factor and covariant scalars of the asymptotic FLRW model $\bar\M$. Evidently, the asymptotic conditions \eqref{FLRW1b} and \eqref{FLRW1c} clearly
identify the variables $\Drho,\,\DDth,\,\DDKK$ as GI exact fluctuations (they vanish as $r\to\infty$) and
$\rho_q,\,\Theta_q$ as GI background variables in an asymptotic FLRW
background. The spatial curvature requires special considerations,
since $\KK_q$ is not GI if the asymptotic FLRW background is spatially
flat ($\bar \KK=0$ holds while $\KK_q$ is in general nonzero for
finite $r$).  

\subsection{Asymptotic non--local exact fluctuations.} 

\begin{figure}
\begin{center}
\includegraphics[scale=0.4]{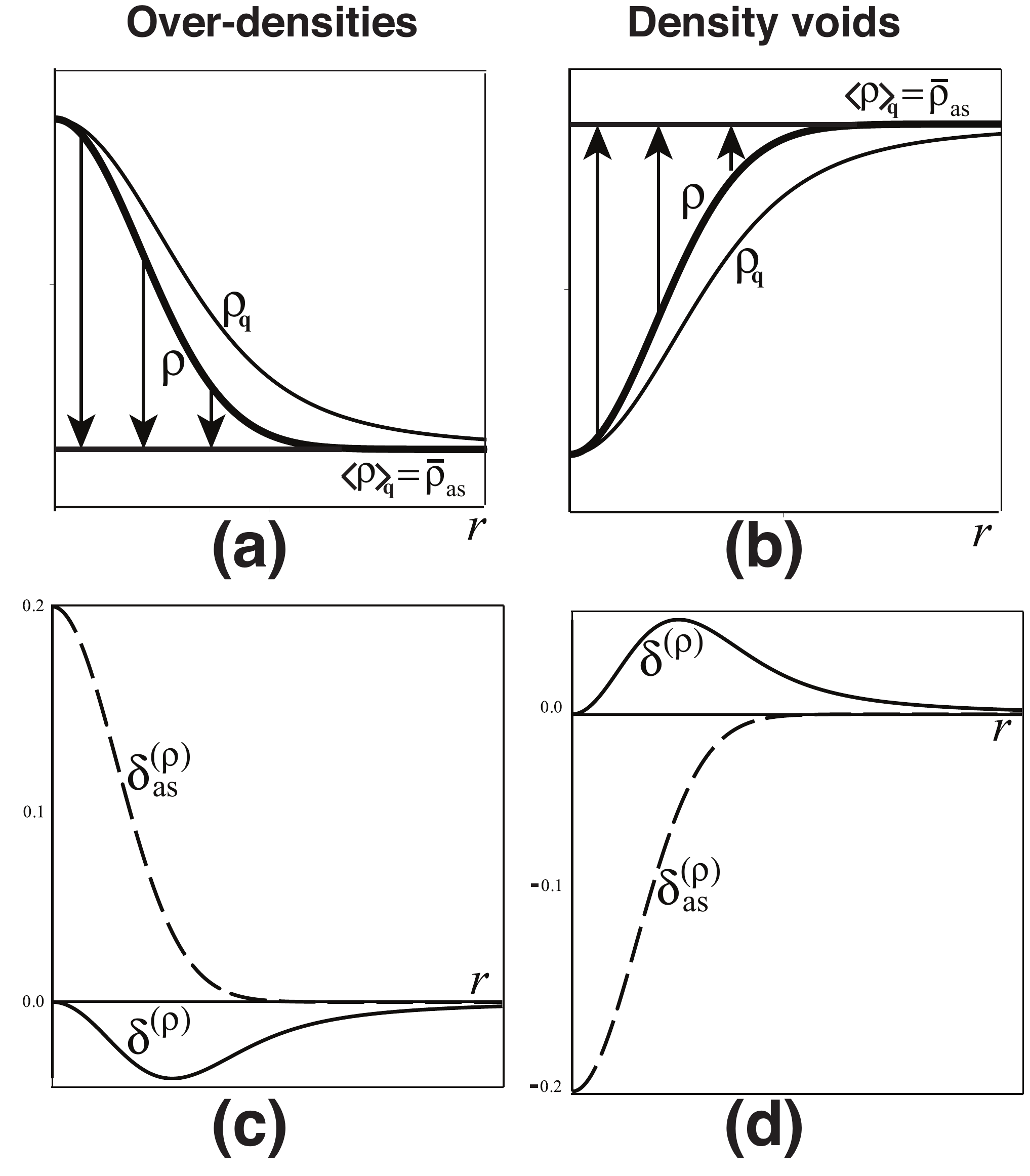}
\caption{{\bf Asymptotic non--local exact density fluctuations.} The panels (a)--(b) describe generic radial density profiles (over--density and density void) of an LTB model that converges radially to a FLRW spacetime as $r\to\infty$, hence the FLRW ``background'' density $\rhobaras$ is given by the average evaluated for a domain comprising the whole time slice. Panels (c) and (d) depict the corresponding exact density fluctuations, which follow by comparing local values of $\rho$ with the global average $\rhobaras$ and is the density contrast with respect to a global density average. Notice that the exact fluctuations vanish as $r\to\infty$.}
\label{FIG3}
\end{center}
\end{figure}

Non--local exact fluctuations can also be defined for asymptotic domains ($\DD[r_b]$ for $0\leq r<r_b$ but $r_b\to\infty$) in LTB models admitting radial convergence to FLRW. These fluctuations are depicted in Figure~\ref{FIG3}. We will denote these non--local fluctuations by the subindex label~ ${}_{\textrm{\tiny{as}}}$ (which stands for ``asymptotic''), as in this case $\Aav_q$ becomes the global asymptotic average of $A$ in the the whole time slice\footnote{The q--average of covariant scalars coincides with their standard average from Buchert's formalism in the radial asymptotic limit of LTB models that converge in this limit to an FLRW
  spacetime, as the back--reaction term vanishes (see proof in
  \cite{sussBR}).}:  
\ba
  \lim_{r_b\to\infty}\rhoav_q[r_b]=\rhobaras,&&\qquad
\lim_{r_b\to\infty}\Thav_q[r_b]=\Thbaras,\notag\\
&&\lim_{r_b\to\infty}\KKav_q[r_b]=\Kbaras,\label{asNL1}
\ea
so that the $\Aav_q$ and the $A_q$ have the same asymptotic limits
given by the asymptotic FLRW scalars $\bar A$, leading to  
\bse
\ba 
\Drhoas\equiv
\lim_{r_b\to\infty}\Drhonl=\frac{\rho-\rhobaras}{\rhobaras},\label{asNL2a}\\ 
\DDthas \equiv \lim_{r_b\to\infty}\DDthnl=\Theta-\Thbaras,\label{asNL2b}\\
\DDKKas \equiv \lim_{r_b\to\infty}\DDKKnl=\KK-\Kbaras,\label{asNL2c}
\ea
\ese
which depend on $t$ and $r$ and are (in general) nonzero for finite
$r$, though (from the limit \eqref{FLRW1d})the fluctuations above do vanish in the limit
$r\to\infty$ (see Figure~\ref{FIG3}). 

It follows readily from \eqref{asNL1} and
\eqref{asNL2a}--\eqref{asNL2c} that the asymptotic exact fluctuations
$\Drhoas,\,\DDthas,\,\DDKKas$ are GI perturbations (they vanish in
$\bar\M$), while the asymptotic q--averages $\rhobaras,\,\Thbaras$ are
the GI background variables ($\Kbaras$ is only a GI variable when the
asymptotic FLRW model is not spatially flat).   

The evolution of non--local q--perturbations for an asymptotic FLRW background can be fully determined by applying \eqref{asNL1} and \eqref{asNL2a}--\eqref{asNL2c} to \eqref{NL1c}--\eqref{NL1d} and to the spatial curvature perturbation constraint in \eqref{constrNL1}--\eqref{constrNL2}, leading to:
\bse\ba 
  \dDrhoas &=& -\left[1+\Drhoas\right]\,\DDthas,\label{NL2c}\\
  \dDDthas &=&-\left(2\Thbaras-\frac{4}{3}\Theta_q+\DDthas\right)\DDthas-\notag\\
&&\,\frac{2}{3}(\Theta_q-\Thbaras)^2-4\pi\rhobaras\,\Drhoas,\label{NL2d}\\
  \DDKKas &=&\frac{8\pi}{3}\rhobaras\,\Drhoas-\frac{2}{9}\Thbaras\DDthas,\label{constr2}
\ea\ese
where $\rhobaras,\,\Thbaras$ and $\Kbaras$ (which are determined by the background subsystem \eqref{NL1a}--\eqref{NL1b}) take the following analytic forms
\beq 
\frac{\Thbaras^2}{9} =\frac{8\pi}{3}(\rhobaras+\Lambda)-\Kbaras,\; \rhobaras=\frac{\rhoibaras}{\abaras^3},\; \Kbaras=\frac{\Kibaras}{\abaras^2},\label{asbackgr1}
\eeq 
with $\rhoibaras=\rhobaras(t_0),\,\Kibaras=\Kbaras(t_0)$. As \eqref{NL1c}--\eqref{NL1d}, this system must be supplemented by \eqref{FFq1}--\eqref{FFq2} to determine $\Theta_q$. As for the non-local fluctuations, we can derive the following second order equation for $\Drhoas$:
\ba  
 && \ddDrhoas -\frac{\left[\dDrhoas\right]^2}{1+\Drhoas}+\left[2\Thbaras-\frac{4}{3}\Theta_q\right]\dDrhoas-\notag \\
&&\;\left[4\pi \rhobaras\Drhoas -2(\Theta_q-\Thbaras)^2\right](1+\Drhoas)=0,
\label{secord3}\ea
\noindent which is equivalent to \eqref{secord2}. In particular, if the asymptotic FLRW background is spatially flat ($\Kbaras=0$) we can use explicit analytic expression for the background variables \eqref{asbackgr1} in \eqref{asNL2c}--\eqref{constr2} and \eqref{secord3}.

We remark that the asymptotic non--local fluctuation $\Drhoas$ provides, for LTB models which radially converge to an asymptotic FLRW background, a covariant and GI description of the density contrast with respect to the asymptotic FLRW background.
\section{Linear regime in LTB models.} 
\label{Linear-Regime}

The exact fluctuations (local and non--local)  that we have introduced provide an exact non--linear measure of the deviation of LTB dynamics with respect to a domain dependent FLRW background. In order to compare these objects with linear perturbations used in the literature we need to define a linear regime involving specific evolution times in which this deviation is also linear.  

Let $\bar A(t)$ (for $A=\rho,\Theta,\,\KK$) be the covariant scalars characterizing an FLRW background on a given domain (bounded through Eq.~\eqref{darmois} or asymptotic). The necessary and sufficient conditions for a linear regime follow by assuming that an arbitrarily small positive number $\epsilon\ll 1$ exists, such that for all $0\leq r\leq r_b$ along a domain $\DD[r_b]$ (all $r$ for asymptotic domains) in a fiducial time slice (say $t=t_0$), the following relations hold
\beq |A_{q0}(r )-\bar A_0|\sim O(\epsilon),\qquad |r A'_{q0}( r)|\sim O(\epsilon),\label{lincond1}\eeq
where $\bar A_0=\bar A(t_0)$ and $O(\epsilon)\ll1$ denotes order $\epsilon$ (or linear first order deviations) for suitable expansions (see Appendix~\ref{app:c}). Since $a_0=\Gamma_0=1$ (choice of radial coordinate),  then \eqref{Drho}--\eqref{DDKK} and \eqref{relperts1}--\eqref{relperts2} together with Eq.~\eqref{lincond1} imply that the following quantities are all $\sim O(\epsilon)$ for all $r$
\bse \ba  
 &&|\Drho(t_0)|,\quad|\DDth(t_0)|,\quad|\DDKK(t_0)|,\label{lincond2a}\\
   &&|\Drhonl(t_0)-\Drho(t_0)|,\quad |\DDthnl(t_0)-\DDth(t_0)|,\notag \\
&&\qquad\qquad|\DDKKnl(t_0)-\DDKK(t_0)|,\label{lincond2b}\\	
   &&\qquad|A_{q0}-A_0|,\quad |\Aav_{q0}-A_0|,\quad \notag \\
&& |A_{q0}-\Aav_{q0}|,\quad |A_0-\bar A_0|,\quad |A'_{q0}-A'_0|,\label{lincond2c}
\ea\ese
As a consequence of  Eq~\eqref{lincond1}, it is also straightforward to show (see proof in Appendix~\ref{app:c}) that a time range containing $t_0$ exists such that the metric variables $a$ and $\Gamma$ in \eqref{ltb2} satisfy at all $r$
\beq a-\bar a\sim O(\epsilon),\qquad \Gamma -1 \sim O(\epsilon),\label{lincond3} \eeq  
and thus, from the scaling laws \eqref{rhoKKHH}--\eqref{DDH}, the relations \eqref{lincond2a}--\eqref{lincond2c} hold for this time range
\bse\ba  
 &&|\Drho|\sim O(\epsilon),\quad|\DDth|\sim O(\epsilon),\quad|\DDKK|\sim O(\epsilon),\quad
\label{lincond4a}\\
   &&\Drhonl \approx \Drho,\quad \DDthnl\approx \DDth,\quad \DDKKnl\approx \DDKK,\label{lincond4b}\\
  && A_q\approx \Aav_q\approx A\approx \bar A,\quad A'_q\approx A'.\label{lincond4c}\ea\ese
which implies that $\dDrho,\,\dDDth$ and $\dDrhonl,\,\dDDthnl$ are $O(\epsilon)$ quantities because of the evolution equations \eqref{FFq1}--\eqref{FFq4} and \ref{NL1a}--\eqref{NL1d}. In fact, we can identify the linear regime in terms of a linear deviation between LTB and FLRW metric functions through Eq.~\eqref{lincond3} and a linear deviation between $A_q,\,\Aav$ and $A$ from the background scalars $\bar A$ through Eqs.~\eqref{lincond2c} and \eqref{lincond4c}. On the other hand,  products of all $O(\epsilon)$ quantities in these evolution equations are of (at least) quadratic order $O(\epsilon^2)\ll O(\epsilon)$, and thus are negligible in the linear regime.

Considering the characteristic features  of  the linear regime, it is
important to emphasise the following points:
\begin{itemize}
\item  The general evolution of LTB models is non--linear. Hence, the
linear regime is only valid for a restricted evolution time range of an
LTB model in which the fluctuations  and relations we presented above
remain of $O(\epsilon)$ (this time range is defined rigorously in Appendix~\ref{app:c}). 
The linear regime is usually defined with respect
to a spatially flat dust FLRW background (Einstein de Sitter of
$\Lambda$--CDM) at initial times after the last scattering surface.
However, it can also be defined with respect to a spatially curved
background.
\item Under a linear regime, the non--linear second order equations for the
density fluctuations  \eqref{secord1}, \eqref{secord2} and \eqref{secord3}
 are reduced to,
\bse\ba \ddDrho
+\frac{2}{3}\Theta_q\dDrho-4\pi\rho_q\Drho=0,\label{secordlin1}\\
\ddDrhonl
+\frac{2}{3}\Thav_q\dDrhonl-4\pi\rhoav_q\Drhonl=0,\label{secordlin2}\\
\ddDrhoas +\frac{2}{3}\Thbaras
\dDrhoas-4\pi\rhobaras\Drhoas=0,\label{secordlin3}
\ea\ese
all of which match (at $O(\epsilon)$) the well known linear evolution
equation for dust perturbations in the synchronous-comoving gauge as discussed in the
next section. As a consequence, these density fluctuations in their
linear regime can be expressed as the linear superposition  $\Drho=
C_{+}( r) D_{+}(t) +  C_{-}( r) D_{-}(t)$ in terms of the growing $(+)$
and decaying density modes $(+)$ .  The explicit analytic form of the
functions $C_\pm,\,D_\pm$ are given in \cite{sussmodes}. In particular, for a spatially flat FLRW
background at early times (so that the effect $\Lambda$ is negligible), we have from equations (27), (36) and (38) of  \cite{sussmodes}  
\beq   
 \Drho  \approx -\frac{2}{5}\left(\Drho_0-\frac{3}{2}\DKK_0\right) \Omega^K_{q0}\, t^{2/3}-\frac{r\tbb'}{t},\label{Drholin}
\eeq
\noindent with $ \Omega^K_{q0}\equiv {\KK_{q0}}/{\HH_{q0}^2}$. This is a solution of  \eqref{secordlin1}--\eqref{secordlin3} (notice that the other fluctuations $\Drhonl$ and $\Drhoas$ take the same form as $\Drho$ at linear order). The generalisation of the linear density modes to the exact non--linear regime (for the case $\Lambda=0$) is discussed extensively in \cite{sussmodes}.

\item It is worthwhile comparing the linear limit of the fluctuations we
have introduced with those obtained by Zibin \cite{zibin} for LTB models
that are ``close'' to an Einstein de Sitter FLRW background ($\Lambda=0$).
The linear limit of $\Drho$ in \eqref{Drholin} is 
formally identical to Zibin's equation (A1) in the Appendix of \cite{zibin}, and 
the linear expansion of the exact growing mode  obtained in \cite{sussmodes} (first term in the right hand side of \eqref{Drholin})  exactly coincides with Zibin's
equation (A3) in \cite{zibin}. The direct relation between this growing
mode linear expansion and the small deviations from spatial flatness expressed in terms of $\Omega_{q0}^K=-\KK_{q0}/\HH_{q0}^2\approx 0$ in \eqref{Drholin} is what Zibin calls ``curvature
fluctuation'' and motivates his comment that  ``{\it the curvature
perturbation consists of just the growing mode}''.  However, this quantity
is not the complete curvature perturbation (see also next section).

\end{itemize}

\noindent
In the following sections we will use the properties of the linear regime
to compare the local and non--local exact fluctuations with perturbations
from other formalisms usually employed in the literature.

\section{Connection to the cosmological perturbation theory.}
\label{sec:CPT}

The linear regime of the LTB exact fluctuations is the right framework to link them with quantities of the more familiar perturbative formalisms. Our focus here is to draw equivalences between exact fluctuations and the standard quantities in the metric based Cosmological Perturbation Theory (CPT) (see e.g. \cite{Zel:70,tomita,Pee80,bardeen} for pioneering work and \cite{malik:wands} for a review of CPT). 
\subsection{Perturbative fluid evolution}

CPT studies cosmological models relying on the principle of overall homogeneity and isotropy, properties that define a background FLRW spacetime with averaged time-dependent elements.  
Inhomogeneities are described by perturbations on this background but, because of the gauge freedom, the relation between these two manifolds is not unique. This means that a perturbative description of an inhomogeneous cosmological spacetime requires a complete gauge specification, so that the perturbations find a physical meaning. To describe the LTB spacetimes of Eq.~\eqref{ltb2}, where the proper time at every point is the cosmic time, it is convenient to use the synchronous-comoving gauge of the CPT formalism. In this gauge the proper time of every observer is the cosmic time.   
The congruence of observers is given by the common four-velocity $u^a= (1,0,0,0)$, which coincides with the unitary normal to the hypersurfaces of constant time. Defined in this way, the four-velocity of observers is comoving and isochronous with the cosmic fluid, and it remains so in time when matter is described by a dust source, which is precisely the case of concern~\cite{tomita,Bru-Ma-Mol} (though $u^a$ may remain comoving and isochronous with nonzero pressure if $\dot u_a=0$, see \cite{suss2009,sussQL}).
%
%

The set of relevant perturbative quantities is given by  the matter density perturbation, the perturbative expansion and the curvature perturbation, defined from departures with respect to the homogeneous parameters $\rhobar(t),\,\Thbar(t)$ and $\Kbar(t)$ of the background, \emph{average} FLRW spacetime: 
\ba  
\drhocpt &=&\frac{\rho(\bx,t) - \bar{\rho}(t)}{\bar{\rho}(t)},\quad 
\Thetacpt=\Theta(\bx,t)- \bar\Theta(t),\notag\\
&&\KKcpt=\KK(\bx, t) -\bar\KK(t). 
\label{CPTdefs}\ea
%

The CPT quantities above are closely related to non--local exact fluctuations \eqref{nl1}--\eqref{nl2}, as the background scalars $\{\bar\rho,\,\bar\Theta,\,\,\bar\KK\}$ can be rigorously identified with the averages $\{\rhoav[r_b],\,\Thav[r_b],\,\KKav[r_b]\}$ along bounded domains $\DD[r_b]$ (if we consider a Swiss Cheese configuration) or asymptotic averages $\{\rhobaras,\,\Thbaras,\,\Kbaras\}$
(if we consider linear perturbations on global domains that correspond to the whole slice in which $r$ is finite but $r_b\to\infty$). Since the linear regime conditions 
\eqref{lincond4c} imply that for every domain $A\approx \Aav_q$ and $\Aav_q\approx A_q$, CPT perturbations are equivalent to linearized ({\it i.e.} first order) non--local or asymptotic exact fluctuations. That is,
\bse
\ba  
 &&\hbox{Swiss Cheese:}\notag\\
 &&\drhocpt{}_1 \approx \Drhonl{}_{\tiny 1}, \quad \Thetacpt{}_{\tiny 1} \approx \DDthnl{}_{\tiny 1},\quad  \KKcpt{}_{\tiny 1} \approx \DDKKnl{}_{\tiny 1},\qquad
\label{equivalence:nl}\\
&&\hbox{Asymptotic FLRW:}\notag\\
 && \drhocpt{}_{\tiny 1} \approx \Drhoas{}_{\tiny 1}, \quad \Thetacpt{}_{\tiny 1} \approx \DDthas{}_{\tiny 1}, \quad \KKcpt{}_{\tiny 1} \approx \DDKKas{}_{\tiny 1},\qquad
 \label{equivalence:as}
\ea 
\ese 
%
where the subindex ${}_1$ denotes first order expansion around background values. These first order forms are governed by precisely the same set of equations following the above correspondences. Namely, 
\bse\ba
   \ddrhocpt{}_1 &=& - {\Thetacpt}_1, \label{CPT:cont1}\\
  {\dThetacpt}{}_{1} &=& - \frac23 \bar{\Theta} \Thetacpt{}_{1} - 4\pi\bar{\rho} \drhocpt{}_1\,, \label{CPT:ray1}\\
   \frac13 \bar{\Theta} \Thetacpt{}_1 &+&\frac32 \KKcpt{}_1= 4 \pi  \bar\rho \drhocpt{}_1 
\,, \label{CPT:const1}
\ea
\ese

\noindent which are the energy conservation equation, the Raychaudhuri equation and the energy constraint (time-time component of the Einstein equations) at linear order (see e.g. \cite{ellis-book} for their derivation). 
These equations correspond to the following linearized form of the evolution equations (bounded domains) of the exact fluctuations \eqref{NL1a}--\eqref{NL1d} and to the exact form given in \eqref{constrNL1}--\eqref{constrNL2}:  
\bse\ba 
  \dDrhonl &=& -\DDthnl,\label{lin1a}\\
  \dDDthnl &=&-\frac{2}{3}\Thav_q\DDthnl-4\pi\rhoav_q\,\Drhonl,\label{lin1b}\\
  \frac{1}{3}\Thav_q\DDthnl&+& \frac32 \DDKKnl =4\pi\rhoav_q\,\Drhonl,\label{lin1c}\ea\ese
or the linear version of equations \eqref{NL2c}--\eqref{NL2d} and the exact form \eqref{constr2} for asymptotic domains,
\bse\ba 
  \dDrhoas &=& -\DDthas,\label{lin2a}\\
  \dDDthas &=&-\frac{2}{3}\Thbaras\DDthas-4\pi\rhobaras\,\Drhoas,\label{lin2b}\\
 \frac{1}{3}\Thbaras\DDthas &+& \frac32 \DDKKas =4\pi\rhobaras\,\Drhoas, \label{lin2c}\ea\ese
where we omitted the subscript ${}_1$ to simplify the notation. The identification of variables in the sets of equations above confirm the equivalences established in Eqs.~\eqref{equivalence:nl} and~\eqref{equivalence:as}. 
Note that the matter density evolution equation, obtained from a combination of the equations in the above system, will thus be equivalent to the non--linear evolution equations \eqref{secord2} and \eqref{secord3} of the exact fluctuations formalism expanded at linear order in \eqref{secordlin2}--\eqref{secordlin3}. 
 
\subsection{Metric elements and the curvature perturbation}
Let us now use the simple $q-$scalars scheme to relate the
metric elements of CPT with the corresponding LTB quantities. To proceed we can compare term by term the linearised version of the LTB metric in  
\eqref{ltb2} with the perturbed FLRW metric in a synchronous and
comoving gauge. Considering exclusively scalar fluctuations we
can write the LTB metric \eqref{ltb2} as 
\beq
  \dd s^2 =-\dd t^2+ a^2(r,t)\left(\delta_{i j} + \left[2 r \frac{a'}{a} +
  \KK_{q0} r^2\right]\nabla_i r \nabla_j r \right) \dd x^i\dd x^j,
\label{ltb3}
\eeq

\noindent while the perturbed FLRW line element is,
\beq 
  ds^2 = - dt^2 + \abar^2(t) \left[(1 - 2 \psi(\bx,t))\delta_{ij} + 2 \nabla_i
  \nabla_j E(\bx,t) \right]dx^idx^j. 
\label{sync:metric}
\eeq
where $i, j$ represent the cartesian coordinates covering the spatial part of both line elements. We can compare these metric elements considering (from Eq.~\eqref{lincond3} and the results of Sec.~\ref{Linear-Regime} and Appedix~\ref{app:c}) that $a \approx \abar$ and $\Gamma\approx 1$ hold for local and non--local exact fluctuations in the linear regime (lowest order in inhomogeneities) and, from Eq.~\eqref{FLRW1a}, asymptotically in all LTB models converging to a spatially flat FLRW background as $r\to\infty$ \cite{RadAs}. We can thus relate the scalar potentials $E$ and $\psi$ in \eqref{sync:metric}  (cf. \cite{wands:2009}) to LTB metric functions at linear order 
\beq
  E (r,t) = \frac12 G(t,\alpha)\,, \quad \mathrm{and} \quad \psi(r,t) = 
\frac{\partial}{\partial \alpha} G(\alpha) - \left(\frac{a- \abar}{\abar}\right).
\label{metric:ltb}
\eeq
where we introduced the variable $\alpha= r^2$ and the function
\beq
  G(t,\alpha) =  \int_0^{\alpha} d\tilde\alpha \int_0^{\tilde \alpha} d\beta \left(\frac{d}{d\beta} \ln(a(t,\beta)) +
\frac{\KK_{q0} (\beta)}{4}\right).
\label{metric:pert}
\eeq 
This function is evaluated at each constant time $t$ at which the linear regime is valid. 

To end this section let us show that the equivalence of metric elements 
significantly simplifies the demonstration that the curvature
perturbation of CPT is preserved over time at linear order (the proof
in the CPT formalism can be found, e.g., in
\cite{Lyth:1984gv}). Considering only perturbations of
scalar nature, we define the comoving curvature perturbation
$\mathcal{R}_c$ from the spatial curvature scalar as 
\beq
\nabla^2{\mathcal{R}_c} \equiv \frac32 {\bar a^2} \KKcpt{}_1, 
\eeq

\noindent where $\nabla^2 = \delta^{ij}\nabla_i\nabla_j$.  In the asymptotically flat FLRW model, we use this definition together with Eqs.~\eqref{rhoKKHH}, \eqref{DrhoKK}, \eqref{DKK} and \eqref{CPTdefs} to obtain      
\beq
\label{curvature:perts}
  \nabla^2 \mathcal{R}_c = \frac32 \left(\frac{\bar{a}}{a}\right)^2 
\KK_{q 0} \left[\left(\frac23 + \DKK_0\right) \Gamma^{-1} -
  \frac23 \right],
\eeq
  
\noindent with $\DKK_0 ={\DDKK_0/\KK_{q0}}.$ Then the right hand side at lowest order is simplified by noting from
\eqref{CPTdefs} that $\KK_{q 0}$ itself is a first order
quantity. Thus, in the linear regime, 
 $\Gamma \approx 1$ and $a(r,t)\approx \bar{a}(t)$ as shown in the previous section \ref{Linear-Regime} (see a rigorous proof in \ref{app:c}). Considering \eqref{DDKK} and $a_0=1,\,R_0=r$, we can thus write 
\ba
\nabla^2 \mathcal{R}_c &=& \frac32  
\KK_{q0} \left( \delta_0^{(\KK)}\right) = \frac32 \DDKK_0  \notag\\
&=&  \frac{r\,\KK'_{q0}}{2}=\frac{3}{2r^3}\int_0^r{\KK_0\,\bar r^3\dd\bar r},
\label{LTB-CPT:curvature}
\ea

\noindent which is constant in time (preserved by the fluid motion). One can alternatively arrive at this expression by computing the three-curvature Ricci scalar $\KK = \RR / 6$ of the metric in
Eq.~\eqref{sync:metric} at first order. This yields the equivalence
\beq
\frac{3}{2}\bar a^2 \KK_{1}= \nabla^2\mathcal{R}_c =  \nabla^2\psi_1,
\label{ricci:first}
\eeq

\noindent where the metric perturbations are taken at first order. Then using the result in
Eqs.~\eqref{metric:ltb} and \eqref{metric:pert} we recover
equivalence \eqref{LTB-CPT:curvature} at lowest order in perturbative expansion.

The amplitude of the contrast $\DKK_0=\DDKK_0/\KK_{q0}$ of equation \eqref{curvature:perts} is not
restricted to be small because it represents the ratio
between a small (near zero) spatial curvature fluctuation and small (near zero)  background curvature, thus the
gradient of the curvature fluctuation can be large even if the
curvature itself is small as equation \eqref{ricci:first} shows. In the context of CPT, at large scales above the Hubble scale,
$\mathcal{R}_c$ coincides with the gauge-invariant curvature
perturbation of uniform density hypersurfaces $\zeta$. The latter is preserved at non-linear order  throughout the evolution  at scales above the horizon \cite{Lyth:1984gv,Wands:2000dp,malik:sasaki}. These properties have also been
studied through the gradient expansion
of perturbation theory \cite{tanaka,rampf:rigo}. Our results are consistent with these findings in  the non-linear regime the curvature perturbation in spherical symmetry.

\section{Connection to the 1+3 covariant perturbation formalism.} 
\label{CGI-Perts}

A formalism of gauge invariant covariant perturbations on an FLRW background (to be denoted by ``GIC perturbations'') was introduced by Ellis and Bruni \cite{ellisbruni89,BDE,1plus3} (see chapter 10.3 of \cite{ellis-book} for a comprehensive discussion), on the basis of a linearization procedure of the exact evolution equations for the comoving spatial gradients of the density and Hubble expansion scalar: 
\beq  
 \Delta_a = \frac{\ell\, \tilde\nabla_a\rho}{\rho},\quad \Z_a = \ell \,\tilde\nabla_a\Theta, \qquad \mathrm{where}\quad \tilde \nabla_a=h_a^b\nabla_b\,,\label{GIC1}\eeq 
\noindent and where the scale factor $\ell$ is defined by the relation $\dot\ell/\ell =\Theta/3$. To compare the GIC formalism with the perturbations introduced in this paper we consider its application to irrotational dust sources ($p=q_a=\dot u_a=\Pi_{ab}=\omega_{ab}=0$), leading to:
\bse\ba 
\dot \Delta_b &=& -\Z_a -\sigma_a^b \Delta_b,\label{GIC2a}\\
\dot\Z_b &=& -\frac{2}{3}\Theta\Z_a -4\pi \rho\Delta_a- \sigma_a^b\Z_b-\ell\,  \tilde\nabla_a(\sigma_{cd}\sigma^{cd}),\qquad\label{GIC2b} 
\ea\ese
together with the constraints
\beq
\Z_a = -\frac32 \ell \tilde{\nabla}_b \sigma^b_a, \qquad \frac{4\pi}{3} \rho \Delta_a  = - \ell \tilde{\nabla}_b E^b_{~a}.
\eeq
In order to obtain a complete system, these equations must be supplemented by the evolution equations for the shear  and electric Weyl tensors $\sigma^a_b$ and $E^a_b$, see e.g. Appendix A in \cite{ellis-book}. 
Also, applying the operator $\ell\,\tilde \nabla_a$ to the Hamiltonian constraint
\beq
\left(\frac{\Theta}{3}\right)^2 = \frac{8\pi}{3}\rho-\KK -2\sigma_{ab}\sigma^{ab}, \eeq
yields the spatial curvature gradient
\beq  
 \frac32\ell\,\tilde \nabla_a \KK = {4\pi}\rho\Delta_a -\Theta\Z_a - \frac32\ell\,\tilde\nabla_a (\sigma_{cd}\sigma^{cd}).\label{GIC3}\eeq

For LTB models we have $\tilde\nabla_a(f)= f'\delta_a^r$ for any scalar $f$, while $\sigma^a_b$ and $E^a_b$ take the forms
\ba  
 \sigma_{ab}&=&\Sigma\,\ee_{ab},\quad\mathrm{with}\; \Sigma=-\frac{\dot \Gamma}{3\Gamma}=-\frac{1}{3}\DDth, \label{shear1}\\
E_{ab}&=&\EE\,\ee_{ab},\quad \mathrm{with}\;\EE = -\frac{4\pi}{3}\rho_q\Drho,\label{shear2} 
\ea
where $\ee_{ab}=h_{ab}-3n_a n_b$, with $n_a=\sqrt{g_{rr}}\,\delta_a^r$ the unit vector normal orthogonal to $u^a$ and to the orbits of SO(3). Hence, \eqref{GIC2a}--\eqref{GIC2b} and \eqref{GIC3} reduce to the following scalar equations 
\bse\ba  
 \dot\Delta &=& -\Z+2\Sigma \Delta,\label{GIC4a}\\
 \dot\Z&=&-\frac{2}{3}\Theta\Z -4\pi\rho\Delta+2\Sigma\Z-6\ell (\Sigma^2)',\label{GIC4b}
 \ea\ese
together with the constraints
\bse\ba
  \Z = - 3 \ell \left(\Sigma' + \frac{3 }{r}\Gamma \Sigma\right)&,&\; 
\frac{4\pi}{3} \rho\Delta = - \ell \left( \EE' + \frac3r \Gamma \EE \right), \qquad\label{GIC4d}
\\ 
\frac32\ell\,\KK' = {8\pi}\rho\Delta &-&\Theta\Z -9\ell\,(\Sigma^2)',\label{GIC4c}
\ea \ese 
where $\Delta\equiv \ell \,\rho'/\rho$ and $\Z\equiv\ell\,\Theta'$ with $\ell=a\Gamma^{1/3}$. This system must be supplemented by Eqs.~(8a)-(8d) of \cite{part1}.

The connection between the 1+3 GIC gradient variables $\Delta,\,\Z$
and $\ell\,\KK'$ and the q--perturbations follows from the fact that
the latter are also related with radial gradients through
\eqref{Drho}--\eqref{DDKK}. Evidently, $\Z$ and $\ell\,\KK'$ are
analogous to the exact fluctuations $\DDth$ and $\DDKK$, while the
``fractional'' density gradient $\Delta$ is analogous to the
fluctuation $\Drho$. This analogy can be further emphasized by
comparing the exact GIC evolution equations
\eqref{GIC4a}--\eqref{GIC4b} with the exact evolution equations
\eqref{FFq3}--\eqref{FFq4}: if we identify $\Delta,\,\Z$ with
$\Drho,\,\DDth$ then \eqref{GIC4a}--\eqref{GIC4b} and
\eqref{FFq3}--\eqref{FFq4} only differ in their non--linear second
order terms $\Sigma\Delta,\,\Sigma\Z$ and $(\Sigma^2)'$, which (from \eqref{shear1}) can be associated with the second order products $\DDth\Drho,\,[\DDth]^2$ and $\DDth[\DDth]'$. The same relation holds between the spatial curvature gradient constraint \eqref{GIC4b} and the analogue of the curvature
constraint \eqref{DDH}. 

Hence, the linearized form of the evolution equations, 
spatial curvature constraint and second order time evolution of the density perturbation of the GIC and exact fluctuations are indeed  fully equivalent if we restrict ourselves to linear terms:  
\bse
\ba
 \dot\Delta = -\Z\quad\quad &\hbox{vs.}&\quad \dDrho =-
\DDth,\label{GIC8a}\\ 
\hskip-2cm   \dot\Z=-\frac{2}{3}\Theta\Z -4\pi\rho\Delta \quad&\hbox{vs.}&\quad\notag \\
\qquad \qquad \qquad \dDDth=-\frac{2}{3}& \Theta_q& \DDth-4\pi\rho_q\Drho,\qquad\label{GIC8b}\\ 
\frac32  \ell\,\KK' = {4\pi}\rho\Delta
-\Theta\Z\quad&\hbox{vs.}&\quad \notag\\ 
\frac32 \DDKK = 4\pi&\rho_q&\Drho - \Theta_q\DDth,\label{GIC8c}\\ 
  \ddot \Delta+\frac{2}{3}\Theta\dot\Delta-4\pi\rho\Delta=0 \quad&\hbox{vs.}&\quad\notag\\
\ddDrho+\frac{2}{3}\Theta_q&\dDrho&-4\pi\rho_q\Drho=0. \label{GIC8d}
\ea\ese
where now $\ell \approx \abar$ and we neglected the quadratic terms $\Sigma\Delta,\,\Sigma\Z$ and $(\Sigma^2)'$ in the GIC equations \eqref{GIC4a}--\eqref{GIC4c} and the quadratic terms $\Drho\DDth,\,[\DDth]^2,\,[\Drho]^2$  and $[\dDrho]^2$  in \eqref{FFq3}--\eqref{FFq4} and \eqref{secord1}, as required in the linear regime for the exact fluctuations (see the previous section).  In fact, \eqref{lincond3} and \eqref{lincond4c} imply that 
$\Delta\approx \Drho,\,\Z\approx \DDth$ and $\ell\KK'\approx \DDKK$ must hold in a linear regime characterized by negligible spatial gradients:  $\rho'\approx \rho'_q,\,\Theta'\approx \Theta'_q$ and $\KK'\approx\KK'_q$, all of which is consistent with the common gradient structure of the GIC perturbations and the  exact fluctuations. 

The equivalence between the non--local (confined and asymptotic) exact fluctuations and the GIC perturbations in the linear regime follows as a straightforward corollary from the results of the previous section, since negligible spatial gradients implies for all scalars and every domain $\DD[r_b]$ that $A_q(r )\approx A_q(r_b)=\Aav_q$ must hold for inner domains $0\leq r\leq r_b$, hence we must have
\beq
  \label{fluctnl:fluctlocal}
(\Drhonl,\,\DDthnl,\,\DDKKnl)\approx (\Drho,\,\Dth,\,\DKK). 
\eeq

\noindent This is reflected also in the fact that evolution equations \eqref{NL1a}--\eqref{NL1d} and \eqref{FFq1}--\eqref{FFq4} only differ in a quadratic term $(\Theta_q-\Thav_q)^2$ and a first order term in the right hand side of equation \eqref{NL1d}. In a linear regime the quadratic term is negligible and $6\Thav_q-4\Theta_q\approx 2\Theta_q$ holds, which makes both systems formally identical at first order (similar remarks apply for asymptotic perturbations).

\section{Summary and conclusions.} 
\label{sec:summary}

We have discussed in detail how the dynamics of LTB models (assuming $\Lambda>0$) can be fully determined by covariant q--scalars and exact fluctuations (local and non--local) that can be constructed from the dynamical quantities. This description can be  characterized  by a precise, covariant and gauge invariant perturbation formalism in which the q--scalars (or the q--averages) define a FLRW background for any given spherical comoving domain. In the asymptotic limit this domain covers whole time slices and the FLRW background can be identified with global q--averages of covariant scalars. 
\begin{table}
\begin{center}
\begin{tabular}{|c|c|c|}
\multicolumn{3}{c}{\large\textbf{Perturbations-to-Fluctuations dictionary}}\\
\hline
\multicolumn{3}{|c|}{\bf CPT}\\
\cline{1-3}
Perturbations& Exact fluctuations & Eqs. in Sec. \ref{sec:CPT}\\
\hline  \hline
$\delta_{1}$ & $\Drhonl\;$ \textsc{\footnotesize Swiss Cheese} & \eqref{equivalence:nl}, \eqref{CPT:cont1} vs. \eqref{lin1a}\\
&$\Drhoas\;$  \textsc{\footnotesize Asymptotic} & \eqref{equivalence:as}, \eqref{CPT:cont1} vs. \eqref{lin2a}\\
$\Theta_1$ & $\DDthnl\;$ \textsc{\footnotesize Swiss Cheese} & \eqref{equivalence:nl}, \eqref{CPT:ray1} vs. \eqref{lin1b}\\
 & $\DDthas\;$ \textsc{\footnotesize Asymptotic} & \eqref{equivalence:as}, \eqref{CPT:ray1} vs. \eqref{lin2b}\\
$\nabla^2 \mathcal{R}_c $ & $\frac32 \DDKKnl{}_{0}\;$ \textsc{\footnotesize Swiss Cheese}  &  \eqref{equivalence:nl}, \eqref{LTB-CPT:curvature}\\
& $\frac32 \DDKKas{}_0\;$ \textsc{\footnotesize Asymptotic}  &  \eqref{equivalence:as}, \eqref{LTB-CPT:curvature}\\
\hline
\multicolumn{3}{|c|}{\bf GIC}\\
\cline{1-3}
Perturbations & Exact fluctuations & Eqs. in Sec. \ref{CGI-Perts}\\
\hline  \hline
$ \Delta_a = {\ell\, \tilde\nabla_a\rho}/{\rho}$ & $\Drho,\, \Drhonl$ & \eqref{GIC8a}, \eqref{GIC8d}, \eqref{fluctnl:fluctlocal}\\
$\Z_a = \ell \,\tilde\nabla_a\Theta$ & $\DDth,\,\DDthnl$ & \eqref{GIC8b}, \eqref{fluctnl:fluctlocal}\\
$\ell\,\nabla_a\KK$ & $\DDKK,\, \DDKKnl$ &  \eqref{GIC8c}, \eqref{fluctnl:fluctlocal}\\
\hline
\end{tabular}
\end{center}
\label{table1}
\noindent \caption{\noindent Dictionary of perturbations-to-fluctuations expressing the Cosmological Perturbation Theory and the Gauge Invariant-Covariant variables in terms of exact fluctuations. The comparison is carried in the linear regime of the exact fluctuations discussed in section \ref{Linear-Regime}. The subindex ${}_0$ indicates evaluation at an arbitrary fiducial hypersurface $t=t_0$, which can be  taken as present cosmic time. The resulting correspondences are not one-to-one because the different definitions of a non-linear exact fluctuation coincide at linear order (cf. \eqref{fluctnl:fluctlocal}).}
\end{table}

We have thoroughly verified the correspondence of exact fluctuations, local and non--local, in the linear regime, to dust perturbations of the GIC and CPT formalisms. Since LTB models are an exact solution of GR, the description of their dynamics in terms of the exact fluctuations should provide valuable information of the non--linear effects that are missed in the perturbative treatment of dust sources. This new information can be appreciated in our demonstration that the spatial curvature perturbation of CPT (a time preserved quantity  at all scales) is directly related (up to linear terms) to a time preserved quantity associated to the spatial curvature of LTB models. This is an important step towards a better understanding of the connection between linear perturbations and the exact non-linear evolution of inhomogeneous sources evolving initially from small fluctuations. Our work goes beyond the historical treatment of these correspondences \cite{Silk:spherical, Morita:1997ep} in that we consider covariant quantities to compare with gauge-invariant perturbations. Also note that our work does not deal with perturbations on top of the exact LTB solution, an important subject of study in itself \cite{zibin,Clarkson:2009sc,Leithes:2014uda}. 

The equivalences presented in Section~\ref{sec:CPT} are
gauge-invariant relations to the familiar variables in cosmological
perturbation theory, which serve to set initial conditions for spherical collapse of non-linear configurations starting from the linear regime. This aspect will be
explored in future work \cite{future:work}. Another important aspect of the identification of exact fluctuations with synchronous-comoving quantities of the cosmological perturbation theory is that one can construct a direct correspondence between the matter variables of this description and those of the Newtonian cosmology (see, e.g. \cite{wands:2009,Bartolo:10,bruni:2013}), and a description of density profiles of initial inhomogeneities \cite{hidalgo:polnarev}. Our method thus provides a direct path to compare fully non-linear Newtonian and relativistic results and the accuracy of the spherical symmetry assumption throughout the evolution of inhomogeneities.

Finally, while our results are still restricted to the spherically symmetric LTB dust models on an FLRW background, they can be extended to more general spacetimes, sources and backgrounds. In particular, the formalism of q--scalars and q--perturbations can be readily extended to the non--spherical dust Szekeres models \cite{sussbol} (even to the cases that are not quasi--spherical).  {In this way it is possible to construct realistic cosmological models which are both exact solutions and represent perturbations of models of lower symmetry.}

Other possible extensions are to consider local rotational symmetry (LRS) spacetimes \cite{LRS}, that include spherically symmetric geometries with a general fluid source, as well as non--spherical exact perturbations on an LTB background \cite{zibin,dunsbyetal}.
{This will be addressed in future work and will shed light on the analysis of the growth of structure on top of strongly non--linear backgrounds \cite{future:work}. For example over--densities such as clusters, or large voids which are both able to generate large curvature and shear. Furthermore, the coupling of density perturbations to vector and tensor modes will be explored and the corrections induced by this coupling, correctly quantified. These studies, complementary to those using higher order perturbation methods, highlight the importance of including general relativistic effects in modelling the late universe, which will be crucial if we are to correctly interpret observational data from future surveys.}
\begin{acknowledgments}
The authors acknowledge support from grant PAPIIT-UNAM IA-101414. \textit{Fluctuaciones no-lineales en cosmolog\'{\i}a relativista} and PAPIIT-UNAM IN-103413-3, \textit{Teor\'ias de Kaluza-Klein, inflaci\'on y perturbaciones gravitacionales}. RS acknowledges support from grant
SEP–-CONACYT 132132. JCH acknowledges partial support from CONACYT, grant 206832, programme \textit{Apoyos Complementarios para la Consolidaci\'on Institucional de Grupos de Investigaci\'on}. 
\end{acknowledgments}

\begin{appendix}

\section{LTB standard metric variables.}
\label{app:a}
The standard metric used in most of the literature (equation (18.16) of \cite{kras2}) is:
\beq \dd s^2=-\dd t^2 +\frac{R'^2}{1 +2E}\,dr^2+R^2(\dd\theta^2+\sin^2\theta \dd\phi^2),\label{usual1}
\eeq

\noindent where $R$ {satisfies}:
\beq
 \dot R^2 = \frac{2M}{R}+2E+\frac{8\pi}{3}\Lambda\,R^2,\label{usual2}
\eeq
where $R=R(t,r),\,E=E( r),\,M=M( r)$. The metric \eqref{ltb2} and Friedman equation  \eqref{Friedman} follow from \eqref{usual1} and \eqref{usual2} by identifying   
\ba 
  a= \frac{R}{R_0},\quad \KK_{q0} = -\frac{2E}{R_0^2},\quad 
\frac{4\pi}{3}\rho_{q0} = \frac{M}{R_0^3},\notag \\
R_0 = r,\quad \Gamma =\frac{r R'}{R}.\label{usual3}		
\ea
The q--scalars $\rho_q,\,\Theta_q,\,\KK_q$ and the exact fluctuations $\Drho$ and $\DKK=\DDKK/\KK_q$ are
\ba  
 \frac{4\pi}{3}\rho_q = \frac{M}{R^3},\qquad \KK_q = -\frac{2E}{R^2},\qquad \Theta_q = \frac{3\dot R}{R},\label{usual4}\\
  1+\Drho=\frac{M'/M}{3R'/R},\quad \frac{2}{3}+\DKK=\frac{E'/E}{3 R'/R},\label{usual5}\ea
so that the fluctuations $\DDth$ and $\DDKK$ can be easily computed from \eqref{DDH}. The scalars $\rho,\,\KK,\,\Theta,$ follow readily from \eqref{Drho})--\eqref{DDKK}. All quantities introduced, calculated and derived in the paper can be thoroughly ``translated'' to the variables $M,\,E,\,R,\,R'$ by substitution of \eqref{usual3}--\eqref{usual5} into the appropriate expressions.          

\section{Darmois matching conditions.} 
\label{app:b}

Let $\M$ be a generic LTB model described by the metric \eqref{ltb2} and $\bar\M$ an FLRW dust spacetime with metric 
\beq 
 \dd s^2 = -\dd t^2 + \abar^2(t)\left[\frac{\dd r^2}{1-\bar\KK_0\,r^2}+r^2\left(\dd\theta^2+\sin^2\theta\,\dd\phi^2\right)\right],\eeq
characterized by the covariant scalars 
\beq  
 \frac{\Thbar^2}{9} =\frac{\dot{\abar}^2}{\abar^2}=\frac{8\pi}{3}(\rhobar+\Lambda)-\Kbar,\qquad \rhobar = \frac{\rhobar_0}{\abar^3},\qquad \Kbar = \frac{\Kbar_0}{\abar^2},
\eeq
where $\bar\KK_0=k_0/H_0^2$ with $k_0=0,\pm 1$ and $H_0$ is the local Hubble constant (because $\abar$ is dimensionless). 

For arbitrary comoving domains $\DD[r_*]$ an FLRW background $\bar\M$ becomes precisely specified by the Darmois matching conditions along a  ``matching interface'' that is common to $\M$ and $\bar \M$: the 3--dimensional timelike surface $\B[r_*]=\B[r_*](t,\theta,\phi)$ generated by comoving observers at the boundary of each comoving domain $\DD[r_*]$. Fulfillment of these conditions implies the continuity of the induced metric $\gamma_{ab}=g_{ab}-\hat n_a \hat n_b$ and the extrinsic curvature $K_{ab}=-\nabla_b \hat n_a$ of $\B[r_*]$, where $\hat n_a=|g_{rr}|^{1/2}\delta_a^a$ is the spacelike unit one--form normal to $\B[r_*]$ oriented towards increasing $r$. Assuming absence of shell crossings ($\Gamma>0$ holds everywhere in $\M$) and computing $\gamma_{ab}$ and $K_{ab}$ at $\B[r_*]$ from the LTB (limit $r\to r_*$ with $r<r_*$) and FLRW (limit $r\to r_*$ with $r>r_*$) sides, we have: 
\beq a_*=a(t,r_*)=\abar( t),\qquad \KK_{q0*}=\KK_{q0}(r_*)=\Kbar_0,\eeq
leading right away to the continuity of the q--scalars $A_q=\rho_q,\,\Theta_q,\,\KK_q$ at $\B[r_*]$: {\it i.e.} conditions \eqref{darmois} (in \eqref{darmois} we dropped the subindex ${}_*$ because $r_*$ is arbitrary and used instead the symbol $[\,\,]_r$ to denote evaluation at arbitrary fixed $r$). 
 
The fulfillment of Darmois matching conditions does not require the continuity of the metric function $\Gamma$ (gradient of $a$ from \eqref{Gamma}) and of the radial gradients $A'_q$ and $A'$ at $\B[r_*]$. We have from \eqref{Drho}--\eqref{DDKK} at $\B[r_*]$:
\beq  A_* = A_{q*}+\frac{r_* A'_{q*}}{3\Gamma_*}=A_{q*} + \frac{1}{a_*^3 r_*^3}\int_0^{r_*}{A'\,a^3 \bar r^3 \dd\bar r}.\label{Agrads}\eeq
Since the Darmois conditions of Eq.~\eqref{darmois} imply $A_{q*}=\bar A(t)$ but not $A'_{q*}=0$ nor $\Gamma_*=1$, they can be fulfilled if $A_*=\bar A+(r_*A'_{q*})/(3\Gamma_*)\ne \bar A(t)$ (and thus $\Drho_*,\,\DDth_*,\,\DDKK_*$ do not vanish). It is possible, however, to demand (together with Darmois conditions) the continuity of the scalars $A$ at $\B[r_*]$ by the extra condition $A'_{q*}=0$, which forces the conditions $\Drho_*=\DDth_*=\DDKK_*=0$ and yields the so--called Swiss Cheese type of models. Notice from the integral in \eqref{Agrads} that the extra condition $A_*=A_{q*}=\bar A(t)$ necessarily implies a change of sign in $A'$ (for this integral to vanish over the integration domain $0\leq \bar r \leq r_*$ the integrand must change sign). This explains the ``troughs'' and ``humps'' in the  profiles of $\rho$ in Swiss Cheese vacuoles. 

 \section{LTB metric variables in the linear regime.} 
\label{app:c} 
 We  prove in this Appendix that the relations \eqref{lincond3} follow readily under the linear regime assumptions \eqref{lincond1}. Using the quadrature of the Friedman equation \eqref{Friedman} we can express the scale factor $a$  as an implicit function of $t$ with the radial dependence mediated by the initial value functions
 \beq t - \tbb  = F(a,A_{q0})=\int_{\xi=0}^{\xi=a}{\frac{\sqrt{\xi} d\xi}{\sqrt{2\mu_{q0}-\KK_{q0}\,\xi+\lambda\,\xi^3}}},\label{quadrature1}\eeq
 where $A_{q0}=\mu_{q0},\KK_{q0},\lambda$ denotes generically the initial value functions $\mu_{q0}\equiv (4\pi/3)\rho_{q0},\,\lambda\equiv (8\pi/3)\Lambda$. Exactly the same quadrature relates the FLRW scale factor $\bar a(t)$ and the equivalent FLRW initial scalars $\bar A_0=\bar\mu_0,\,\Kbar_0,\,\lambda$ (which are constants): 
 \beq 
t -  \btbb  = F(\abar,\bar A_0)=\int_{\xi=0}^{\xi=\abar}{\frac{\sqrt{\xi} d\xi}{\sqrt{2\bar \mu_{0}-\Kbar_{0}\,\xi+\lambda\,\xi^3}}}.\label{quadrature2}
\eeq
 Combining  \eqref{quadrature1} and \eqref{quadrature2} yields
 \beq  F(a,A_{q0})- F(\abar,\bar A_0)=F_0(A_{q0})- F_0(\bar A_0),\label{quadrature3}\eeq 
 where $F_0(A_{q0}) =F(1,A_{q0}),\,\, F_0(\bar A_0) =F(1,\bar A_0)$ and we have eliminated $\tbb$ and $\btbb$ from the relations $\tbb = t_0-F_0(A_{q0})$ and  $\btbb = t_0-F_0(\bar A_0)$ that follow from  
 the choice of radial coordinate so that $a(t_0,r)=1$ and $\abar (t_0)=1$. Since $F$ is assumed to be analytic and continuous in the functional domain $a,\,A_{q0}$ we can always expand 
\eqref{quadrature3} around $\abar,\,\bar A_0$, which is a point in this domain.  After some algebraic manipulation this expansion yields at first order 
 \beq a-\abar   \approx  F_1(\abar,\bar A_0)(\mu_{q0}-\bar\mu_0)+F_2(\abar,\bar A_0)(\KK_{q0}-\Kbar_0),\label{linexp1}\eeq
 where 
 \ba  
 F_1 \equiv -\abar  \frac{\Thbar}{3}\left[ \frac{\partial (F-F_0)}{\partial\mu_{q0}}\right]_{a=\abar,\,A_{q0}=\bar A_0}, \notag \\
\quad F_2 \equiv -\abar \frac{\Thbar}{3}\left[ \frac{\partial (F-F_0)}{\partial\KK_{q0}}\right]_{a=\abar ,\,A_{q0}=\bar A_0},  \label{F1F2}
\ea 
and we used the fact that $[\partial F/\partial a]_{a=\abar} =\abar \Thbar/3$ with $\Thbar/3=\dot{\abar}/\abar$.  Assuming the conditions \eqref{lincond1} for a linear regime and considering that $F_1$ and $F_2$ in \eqref{F1F2} are functions of $t$ (through $\abar$), the expansion \eqref{linexp1} leads directly to \eqref{lincond3} in the time range satisfying  
\ba 
  F_1(\abar,\bar A_0)(\mu_{q0}-\bar\mu_0)\sim O(\epsilon), \qquad \notag\\
\qquad  F_2(\abar,\bar A_0)(\KK_{q0}-\Kbar_0)\sim O(\epsilon),\label{linexp2}\ea 
which clearly depends on the forms of $F_1$ and $F_2$.  Since $\mu_{q0}-\bar\mu_0$ and $\KK_{q0}-\Kbar_0$ are (assumed) of $O(\epsilon)$ and $a_0-\abar_0=0$, then there is always a range of time sufficiently close to $t_0$ where $a-\abar \sim O(\epsilon)$ holds. In fact,  $a-\abar \sim O(\epsilon)$ as long as $F_1,\,F_2$ are up to $O(1)$.

The linear conditions \eqref{lincond3} on $\Gamma$ follows readily by applying the definition \eqref{Gamma} to \eqref{linexp1}:
\ba  
  \Gamma &=& 1+\frac{ r a'}{a} \approx 1+ \frac{r\mu'_{q0} F_1+r\KK'_{q0} F_2}{\abar +F_1(\mu_{q0}-\bar\mu_0)+F_2(\KK_{q0}-\Kbar_0)}\nonumber\\
  &\approx&  1+r\mu'_{q0}\frac{F_1}{\abar} + r\KK'_{q0}\frac{F_2}{\abar } \notag \\
&\approx& 1+3\mu_{q0}\Drho_0\frac{F_1}{\abar}+3\DDKK_0\frac{F_2}{\abar}, \label{linexp3}\ea
which (since we assume $\mu'_{q0}$ and $\KK'_{q0}$ to be $O(\epsilon)$) has the form $1+O(\epsilon)$ in roughly the same time range as $a-\abar \sim O(\epsilon)$.

\end{appendix}

\end{document}